\documentclass[preprint,aps,preprintnumbers,amsmath,amssymb,nofootinbib,epsf,epsfig]
{revtex4}
\usepackage[utf8]{inputenc}
\usepackage{graphicx}
\usepackage{amsmath}
\usepackage{amssymb}
\usepackage{array}
\usepackage{float}
\usepackage{xcolor}
\usepackage{booktabs}
\usepackage{subcaption}
\usepackage{multirow}
\usepackage{color}
\usepackage{caption}
\usepackage{subcaption}
\usepackage{natbib}
\usepackage[colorlinks, citecolor=cyan]{hyperref}
\usepackage{epsfig}
\usepackage{slashed}
\usepackage{appendix}

\begin{document}
%%%%%%%%%%%%%%%%%%%%%%%%%%%%%%%%%%%%%%%%%%%%%%%%%%%%%%%%%%%%%%%%%%%%%%%%

\newcommand{\be}{\begin{equation}}
\newcommand{\ee}{\end{equation}}
\newcommand{\bea}{\begin{eqnarray}}
\newcommand{\eea}{\end{eqnarray}}
\newcommand{\nn}{\nonumber}
\def\ds{\displaystyle}
\def\s1{\hat s}
\def\para{\parallel}
\newcommand{\mrm}[1]{\mathrm{#1}}
\newcommand{\mc}[1]{\mathcal{#1}}
\def\CP{{\it CP}~}
\def\cp{{\it CP}}
\def\ml{m_\mu}
\title{\large Binwise exploration of vector couplings in  $B_s \to D_s^{(*)} \tau \bar \nu_\tau$ decays}

\author{Ajay Kumar Yadav}
\email{yadavajaykumar286@gmail.com}
\author{Suchismita Sahoo}
\email{suchismita8792@gmail.com}
\affiliation{ Department of Physics, Central University of Karnataka, Kalaburagi-585367, India}
\begin{abstract}
Recent results from the LHCb experiment have confirmed that lepton flavor universality is upheld in flavor changing neutral current processes, such as $B \to K^{(*)} l^+ l^-$.  However, discrepancies remain in the charged current sector, raising questions about the universality of lepton flavors in these processes. To explore this issue, we investigate the decays $B_s \to D_s^{(*)} \tau \bar \nu_\tau$ in different $q^2$ bins, which involve the $b \to c \tau \bar \nu_\tau$ transition. We employ a model independent approach to analyze potential new physics by fitting both real and complex (axial)vector coefficients to the existing $b \to (u, c) \tau \bar \nu_\tau$ data. Our analysis enables us to calculate the branching ratios and angular distributions for the $B_s \to D_s^{(*)} \tau \bar \nu_\tau$ decays in four different $q^2$ bins. Additionally, we evaluate whether lepton flavor universality is maintained in these charged current decays or if deviations suggest the presence of new physics.
\end{abstract}
\maketitle

%==========================================
\section{Introduction}
%=================================================

The Standard Model (SM) is successful in many predictions but falls short in explaining matter-antimatter asymmetry, neutrino mass, dark matter, dark energy, and other anomalies. Recent deviations in semileptonic B meson decays from the SM expectations suggest potential new physics (NP), as observed in measurements of $R_{D^{(*)}}$, $R_{J/\psi}$, and the $\tau$ polarization asymmetry \cite{BUSKULIC1997373,CLEO:1998qvx,HFLAV:2019otj,Bernlochner:2021vlv,Blanke:2018yud,Fedele:2022iib,Dutta:2013qaa, BaBar:2012obs, BaBar:2023kug, Belle:2015qfa,Belle:2017ilt,Belle:2019rba,Belle-II:2025yjp, LHCb:2015gmp,LHCb:2017rln, HeavyFlavorAveragingGroupHFLAV:2024ctg, HFLAV2025, LHCb:2017vlu, Harrison:2020nrv}. 
 Experimentally, the BaBar \cite{BaBar:2012obs, BaBar:2023kug}, Belle \cite{Belle:2015qfa,Belle:2017ilt,Belle:2019rba,Belle-II:2025yjp}, and LHCb \cite{LHCb:2015gmp,LHCb:2017rln, LHCb:2017vlu} collaborations have measured the ratios of branching fractions as
\begin{equation}
     R_{D^{(*)}}=\frac{BR(B\to D^{(*)}\tau\Bar{\nu}_{\tau})}{BR(B\to D^{(*)}l\Bar{\nu}_{l})},~~~~~~~\\
     R_{J/\Psi}=\frac{BR(B_c\to J/\Psi\tau\Bar{\nu}_{\tau})}{BR(B_c\to J/\Psi l\Bar{\nu}_{l})}\,,~~~{\rm where}~ l=e,\mu\,.
\end{equation}
The Heavy Flavor Averaging Group (HFLAV)  reports the average values \cite{HFLAV2025}

\begin{align}
 R_D^{\text{avg}}=&0.347\pm0.025,~~~~~~~~~~~~~~~
    R_{D^{*}}^{\text{avg}}=0.288\pm0.012,
     \end{align} 
which exceed the SM predictions 

   \begin{align}
    R_D^{\text{SM}}=&0.296\pm0.004,~~~~~~~~~~~~~~~
    R_{D^{*}}^{\text{SM}}=0.254\pm0.005\,,
\end{align}
by $1.9\sigma$ and $2.7\sigma$ respectively. Likewise, the ratio $R_{J/\Psi}$, measured by the LHCb experiment \cite{LHCb:2017vlu} and the CMS collaboration \cite{CMS_Collaboration2023,CMS_Collaboration2024}, has been averaged by the HFLAV group \cite{Iguro:2024hyk} as   
\begin{equation}
    R_{J/\Psi}=0.61\pm0.18\,,~~~~~~~
    R_{J/\Psi}^{\text{SM}}=0.258\pm0.004\,,
\end{equation}
which shows a disagreement of $1.9\sigma$ \cite{Harrison:2020nrv} from the SM prediction at the $95\%$ confidence level. This discrepancy suggests a potential violation of lepton flavor universality (LFU) between the tau lepton and the lighter leptons. Semileptonic $B$ decays involving $ b \to c \tau \bar{\nu}_\tau $ flavor-changing charged currents (FCCC) are key for testing SM predictions, exploring potential lepton flavor universality violations (LFUV) and investigating NP beyond the SM.

The LHCb collaboration has reported the first measurements of the branching ratios of the decays $B_s^0 \to D_s^{(*)-} \mu^+ \nu_\mu$~\cite{LHCb:2020cyw}, incorporating external inputs from $BR(B^0 \to D^{(*)-} \mu^+ \nu_\mu)$ as given in Ref.~\cite{ParticleDataGroup:2018ovx}. The measured values are
\begin{eqnarray}
    BR(B_s^0 \to D_s^- \mu^+ \nu_{\mu}) &=& (2.49 \pm 0.12~\text{(stat)} \pm 0.14~\text{(syst)} \pm 0.16~\text{(ext)}) \times 10^{-2}\,, \nonumber \\
    BR(B_s^0 \to D_s^{*-} \mu^+ \nu_{\mu}) &=& (5.38 \pm 0.25~\text{(stat)} \pm 0.46~\text{(syst)} \pm 0.30~\text{(ext)}) \times 10^{-2}\,,
\end{eqnarray}
where the uncertainties are statistical, systematic, and due to external inputs, respectively. The ratio of branching fractions between the $B_s^0 \to D_s^{-} \mu^+ \nu_\mu$ and $B_s^0 \to D_s^{*-} \mu^+ \nu_\mu$ decay modes is determined to be
\begin{equation}
    \frac{BR(B_s^0 \to D_s^- \mu^+ \nu_{\mu})}{BR(B_s^0 \to D_s^{*-} \mu^+ \nu_{\mu})} = 0.464 \pm 0.013~\text{(stat)} \pm 0.043~\text{(syst)}\,.
\end{equation}
Furthermore, the ratios of the exclusive $B_s^0 \to D_s^{(*)-} \mu^+ \nu_\mu$ branching fractions to their $B^0$ counterparts are measured as
\begin{eqnarray}
    R &\equiv& \frac{BR(B_s^0 \to D_s^- \mu^+ \nu_\mu)}{BR(B^0 \to D^- \mu^+ \nu_\mu)} = 1.09 \pm 0.05~\text{(stat)} \pm 0.06~\text{(syst)} \pm 0.05~\text{(ext)}\,, \\
    R^\ast &\equiv& \frac{BR(B_s^0 \to D_s^{*-} \mu^+ \nu_\mu)}{BR(B^0 \to D^{*-} \mu^+ \nu_\mu)} = 1.06 \pm 0.05~\text{(stat)} \pm 0.07~\text{(syst)} \pm 0.05~\text{(ext)}\,.
\end{eqnarray}
These measurements highlight that semileptonic $B_s$ decays exhibit behavior very similar to the well studied $B \to D^{(*)} \ell \nu_\ell$ channels.  Since both $B$ and $B_s$ mesons decay via the same underlying quark level transition $b \to c \ell \nu$,  but differ in spectator quark content, the comparison between them offers a valuable test of hadronic form factor modeling and potential flavor specific new physics contributions. In particular, the consistency of ratios like $R$ and $R^*$ with unity supports the universality of form factors across $B$ and $B_s$ mesons, thereby lending credibility to theoretical extrapolations. Given the well known anomalies in $R_{D^{(*)}}$ and $R_{J/\psi}$, which suggest a possible violation of lepton flavor universality, it is timely to extend similar tests to the $B_s$ meson.  Moreover, semileptonic $B_s$ decays can play an essential role in refining the extraction of the CKM matrix element $|V_{cb}|$, especially when combined with their $B$ meson counterparts. Although no experimental measurements of $R_{D_s^{(*)}}$  are available at present, several theoretical analyses, both model dependent and model independent, have explored these channels~\cite{Koponen:2007fe,Li:2009wq,Bhol_2014,Bordone:2020gao,Dutta:2018jxz,Sahoo:2019hbu,Zhang:2022opp,Sahoo:2021wyc,Blossier:2021xvl,Sahoo:2020wnk,Gubernari:2023rfu,Rahmani:2024pko}. Thus, pursuing the study of $B_s$ semileptonic decays is a well justified and necessary step towards understanding the origin of the observed LFU anomalies.

This study aims to perform a binwise analysis of the decay processes $B_s \to D_s^{(*)} \tau \bar{\nu}_{\tau}$, which include the $b \to c \tau \bar{\nu}_{\tau}$ quark level transition, in a model independent approach by extending the operator structure of the Lagrangian beyond the SM. In this formalism, we identify additional (axial)vector Wilson coefficients that contribute to the SM coefficients. By performing a $\chi^2$ fit to $b \to (u, c) \tau \bar{\nu}_\tau$ data, we determine the best-fit values for these new real and complex parameters. We estimate the binwise branching ratios, forward-backward asymmetries, lepton non-universality (LNU) ratios, and the polarization asymmetries of $\tau$ and $D^*_{s}$ for $B_s \to D_s^{(*)} \tau \bar{\nu}_{\tau}$ decay modes with both real and complex new coefficients. A binwise analysis is particularly sensitive to NP effects, as it preserves localized shape distortions in the $q^2$ distribution that integrated observables may average out. This approach facilitates the disentanglement of different NP Lorentz structures, as emphasized in Refs.~\cite{Sakaki:2014sea, Bhattacharya:2015ida, Celis:2016azn}.

The paper is organized as follows. Section II presents the effective Hamiltonian for semileptonic decays involving the quark level transition $b \to c \tau \bar{\nu}_\tau$ and details the global fit of both real and complex (axial)vector Wilson coefficients using $b \to (u, c) \tau \bar{\nu}_{\tau}$ experimental data. Section III provides detailed expressions for the branching ratios and various angular observables of $B_s \to D_s^{(*)} \tau \bar{\nu}_{\tau}$ decays and includes the numerical evaluation of $B_s \to D_s^{(*)} \tau \bar{\nu}_{\tau}$ decay modes with the constrained new parameters. Finally, Section IV summarizes our results.

%%%%%%%%%%%%%%%%%%%%%%%%%%%%%%%%%%%%%%%
\section{Theoretical Model Formulation}
%=================================================================
\subsection{Effective Hamiltonian}
%============================================================================
The effective Hamiltonian responsible for the $b \to c \tau \bar{\nu}_\tau$ transitions, including only the (axial)vector operator structure extension, is given by \cite{Tanaka:2012nw}
\begin{equation} \label{ham-bclnu}
\mathcal{H}_{\rm eff} = \frac{4G_F}{\sqrt{2}} V_{cb} \left[ \left(1 + V_L \right) \mathcal{O}_{V_L} + V_R \mathcal{O}_{V_R} \right],
\end{equation}
where $G_F$ is the Fermi constant, $V_{cb}$ is the CKM matrix element, and $V_{L,R}$ are the Wilson coefficients, which are zero in the SM and can arise only in the presence of new physics. The corresponding dimension-six effective operators $(\mathcal{O}_{V_{L,R}})$ are given by
\begin{align*}
\mathcal{O}_{V_L} = \left(\bar{c}_L \gamma^\mu b_L \right) \left(\bar{\tau}_L \gamma_\mu \nu_{l L} \right), ~~~~~~~~
\mathcal{O}_{V_R}= \left(\bar{c}_R \gamma^\mu b_R \right) \left(\bar{\tau}_L \gamma_\mu \nu_{l L} \right),
\end{align*}
where $f_{L(R)} = P_{L(R)} f $ are the chiral fermion $f$ fields with $P_{L(R)} = (1\mp\gamma_5)/2$ being the projection operators.

%================================================
%\subsection{New Physics Scenarios}
%====================================================

%==============================================================
\subsection{Numerical Fitting of Model Parameters}
%============================================================
In our analysis, we assume that the same NP operators contribute to both \(b \to c \ell \nu\) and \(b \to u \ell \nu\) transitions, with identical Wilson coefficients. This alignment is supported by phenomenological considerations, such as global analyses using sum rules~\cite{Duan:2024ayo} as well as SMEFT constructions with minimally broken flavor symmetries like \(U(2)^5\)~\cite{Faroughy:2020ina, Greljo:2022cah}, which naturally predict similar NP contributions in third generation dominated processes. Under this assumption, we perform a global fit of the new coefficients to the \(b \to (u,c) \tau \bar{\nu}_\tau\) data. The $\chi^2$ function is defined as 
\begin{equation}
\chi^2(V_{L,R}) = \sum_i \frac{(\mathcal{O}_i^{\rm th}(V_{L,R}) - \mathcal{O}_i^{\rm Expt})^2}{(\Delta \mathcal{O}_i^{\rm Expt})^2 + (\Delta \mathcal{O}_i^{\rm SM})^2}\,,
\end{equation}
where $\mathcal{O}_i^{\rm th}(V_{L,R})$ denotes the theoretical predictions of the observables, $\mathcal{O}_i^{\rm Expt}$ represents the corresponding experimental central values, and $\Delta \mathcal{O}_i^{\rm Expt}$ and $\Delta \mathcal{O}_i^{\rm SM}$ are the experimental and SM uncertainties, respectively.  To scrutinize the signature of new (axial)vector coefficients, we analyzed several possible combinations of both real and complex Wilson coefficients. The details are provided in Table \ref{Tab:scenarios}\,.

%%%%%%%%%%%%%%%
\begin{table}[htp]
\setlength{\tabcolsep}{6pt}
\centering
\begin{tabular}{|l l | l l |l l|}
\hline
\multicolumn{2}{|c|}{\textbf{1D Scenario}} & \multicolumn{2}{|c|}{\textbf{2D Scenario}} & \multicolumn{2}{|c|}{\textbf{4D Scenario}} \\ \hline
Scenario & Coefficient & Scenario & Coefficient & Scenario & Coefficient  \\
\hline
1DS-I &  $Re[V_L]$ & 2DS-I & $(Re[V_L], Re[V_R])$ & 4DS-I & $(Re[V_L], Im[V_L],$ \\
1DS-II &  $Re[V_R]$ & 2DS-II & $(Re[V_L], Im[V_L])$ & & $Re[V_R], Im[V_R])$\\
 &  & 2DS-III & $(Re[V_R], Im[V_R])$ & & \\
\hline
\end{tabular}
\caption{Scenarios of new real and complex (axial)vector coefficients.}
\label{Tab:scenarios}       % Give a unique label
\end{table}
%%%%%%%%%%%%%%%%%%%%%%%%%%%%%
We explored six possible scenarios: two one-dimensional scenarios (1DS-I and 1DS-II), three two-dimensional scenarios (2DS-I, 2DS-II, and 2DS-III), and one four-dimensional scenario (4DS-I), fitting them to the observables of $b \to c \tau \bar{\nu}_\tau$ and $b \to u \tau \bar{\nu}_\tau$ in two different ways:

\begin{itemize} \renewcommand{\labelitemi}{\textbf{--}} \item Case \textbf{A}: Using only the observables associated with the $b \to c \tau \bar{\nu}_\tau$ data. \item Case \textbf{B}: Incorporating both $b \to c \tau \bar{\nu}_\tau$ and $b \to u \tau \bar{\nu}_\tau$ observables. \end{itemize}

For the $b \to c \tau \bar{\nu}_\tau$ observables, we utilized existing data on $R_{D^{(*)}}$, $R_{J/\psi}$, $P_\tau^{D^*}$, $F_L^{D^*}$, and the branching ratio for $B_c \to \tau \bar{\nu}_\tau$, which is estimated to be less than $30\%$ based on the lifetime of the $B_c$ meson. Additionally, we fitted the observables $R_\pi^\ell (= \Gamma(B_u^- \to \tau^- \bar \nu)/\Gamma(B^0 \to \pi^+ l^- \bar \nu_l))$, ${\rm BR}(B_u \to \tau \bar{\nu}_\tau)$, and ${\rm BR}(B^0 \to \pi^+ \tau \bar{\nu}_\tau)$ related to the $b \to u \tau \bar{\nu}_\tau$ transition using the new parameters. Details of these observables can be found in Table \ref{Tab:input-fit}.
%%%%%%%%%%%%%%%%%%%%%%%%%%%
\begin{table}[htb]
\centering
\begin{tabular}{|c|c|c|}
\hline
Observables &~Experimental values~~&~SM Predictions~ \\
\hline
\hline
$R_D$ & $0.347\pm 0.025$ \cite{HFLAV2025} & $0.296\pm 0.004$ \cite{HFLAV2025}\\
$R_{D^*}$ &  $0.288 \pm 0.012$ \cite{HFLAV2025} &  $0.254\pm 0.005$  \cite{HFLAV2025}\\
$R_{J/\psi}$  &  $0.61\pm 0.18$   \cite{LHCb:2017vlu,CMS_Collaboration2023,CMS_Collaboration2024,Iguro:2024hyk} & $0.258\pm 0.004$  \cite{Harrison:2020nrv} \\
$P_{\tau}^{D^*}$  & $-0.38^{+0.53}_{-0.55}$ \cite{Iguro:2024hyk}  & $-0.497\pm 0.007$ \cite{Iguro:2020cpg}\\
$F_{L}^{D^*}$  &  $0.49\pm 0.05$ \cite{Belle:2019ewo,Chen:2868260,LHCb:2023ssl,Iguro:2024hyk} & $0.464 \pm 0.003$ \cite{Iguro:2020cpg} \\
${\rm Br}(B_c \to \tau \bar \nu_\tau)$ & $<30\%$  \cite{Alonso:2016oyd, Li:2016vvp, Celis:2016azn} & $(2.29 \pm 0.09) \times 10^{-2}$ \cite{Zuo:2023dzn,Iguro:2024hyk}\\
\hline
$R_\pi^l$  & $0.73 \pm 0.14$ \cite{Iguro:2023prq, Belle-II:2018jsg} & $0.54\pm 0.04$ \cite{Iguro:2023prq, Tanaka:2016ijq}\\
${\rm BR}(B_u \to \tau \bar \nu_\tau )$ & $ (1.09\pm 0.24) \times 10^{-4}$ \cite{ParticleDataGroup:2024cfk}& $(0.87\pm0.05)  \times 10^{-4}$  \cite{Zuo:2023dzn,Iguro:2024hyk} \\
${\rm BR}(B^0 \to \pi^+ \tau \bar \nu_\tau) $  & $< 2.5 \times 10^{-4}$ \cite{Belle:2015qal, ParticleDataGroup:2024cfk}  & $(9.35 \pm 0.38) \times 10^{-5}$ \cite{Du:2015tda}\\
\hline
\end{tabular}
\caption{Observed and theoretical values of quantities employed in the fitting process.} \label{Tab:input-fit}
\end{table}

%%%%%%%%%%%%%%==================

\begin{figure}[htp]
    \centering
    \includegraphics[width=0.42\linewidth]{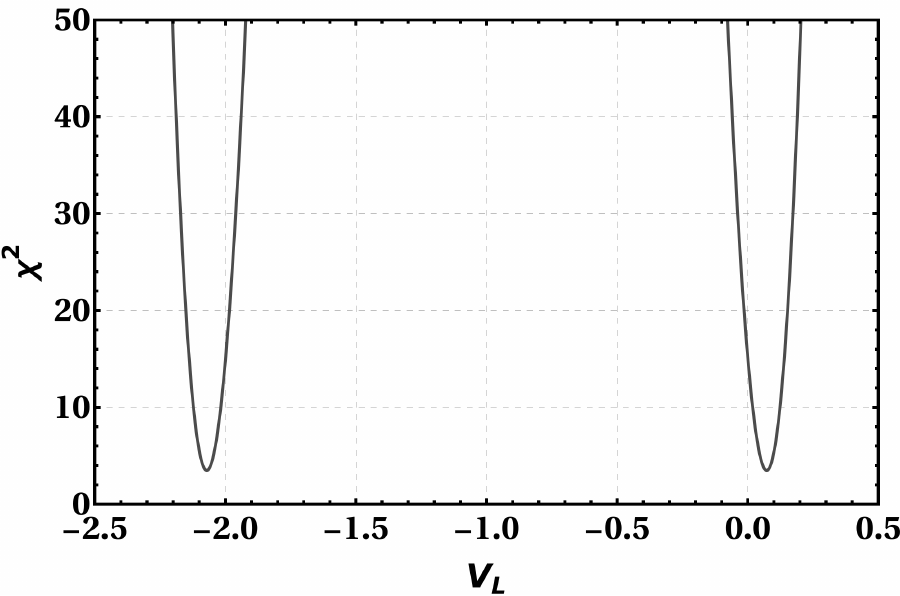} \quad
    \includegraphics[width=0.42\linewidth]{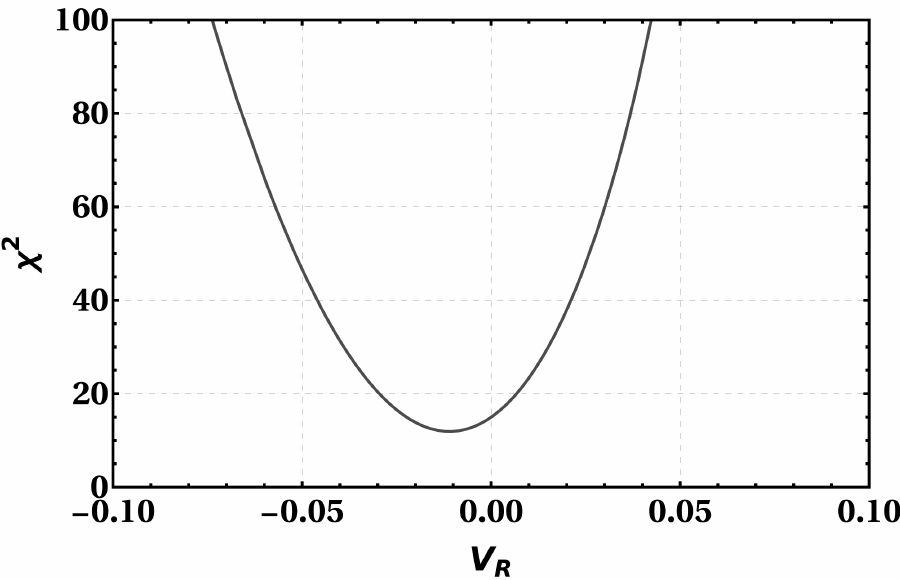}\\
    \includegraphics[width=0.38\linewidth]{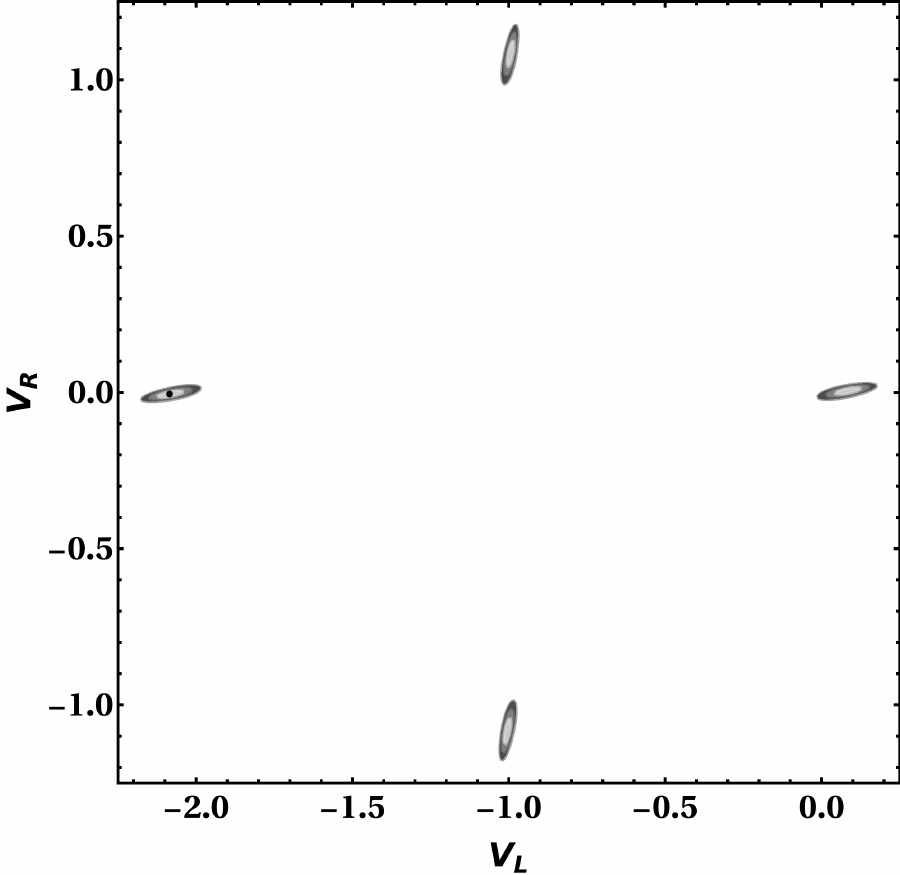}\quad
     \includegraphics[width=0.38\linewidth]{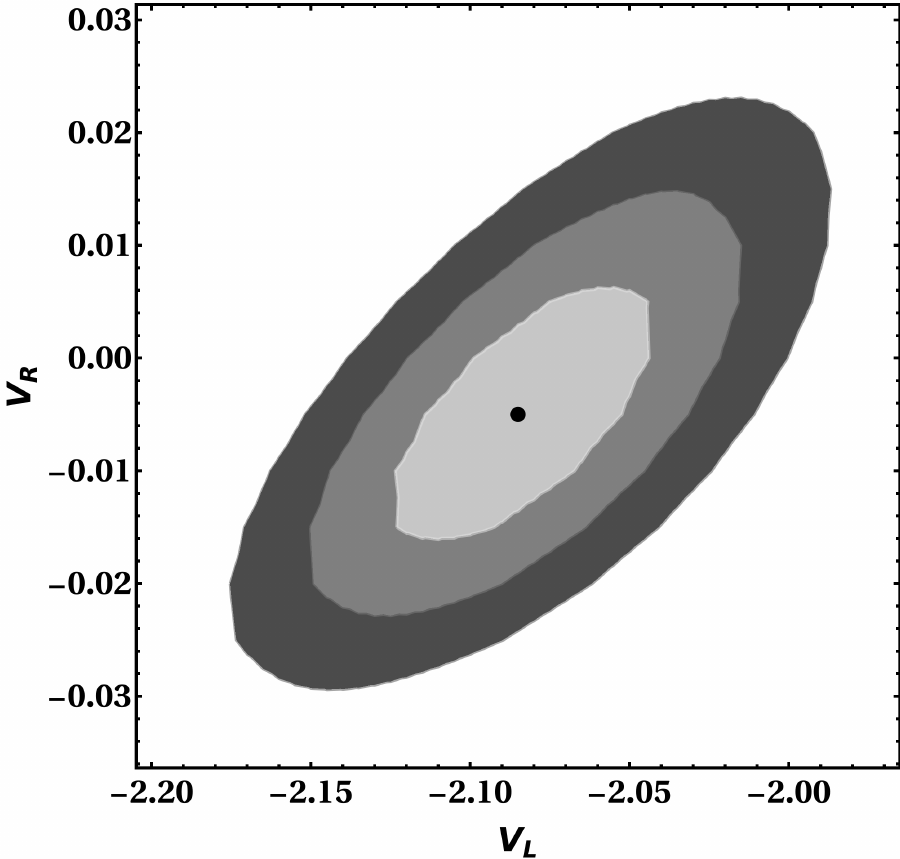}\\
      \includegraphics[width=0.38\linewidth]{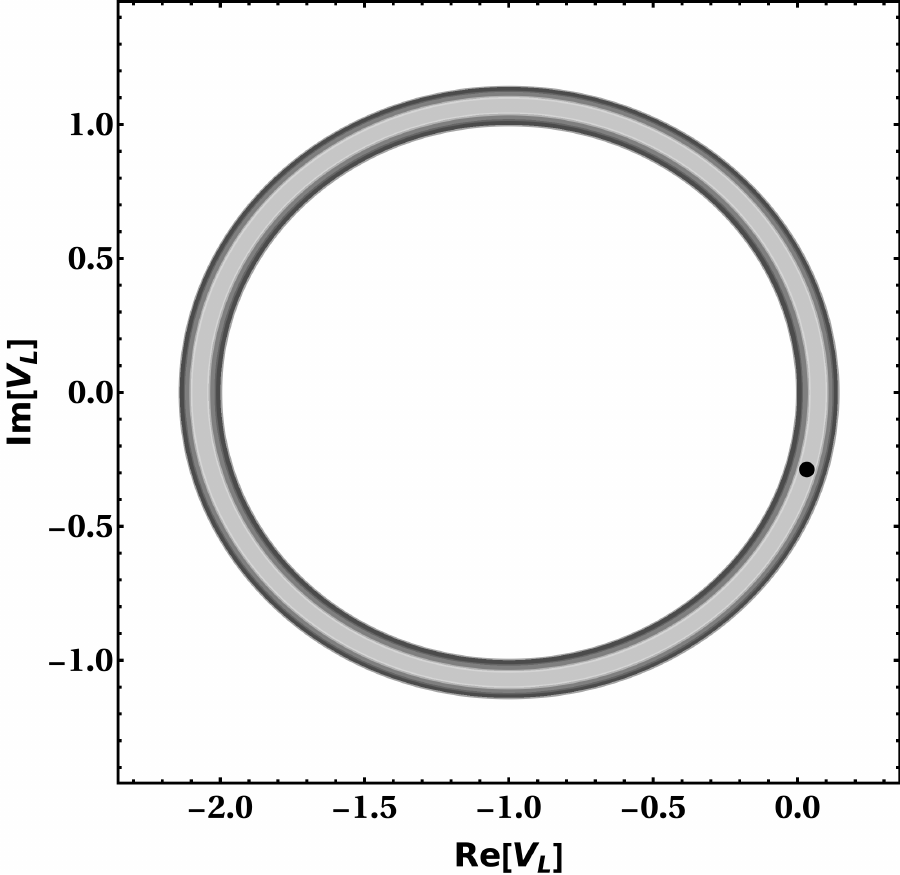}\quad
       \includegraphics[width=0.38\linewidth]{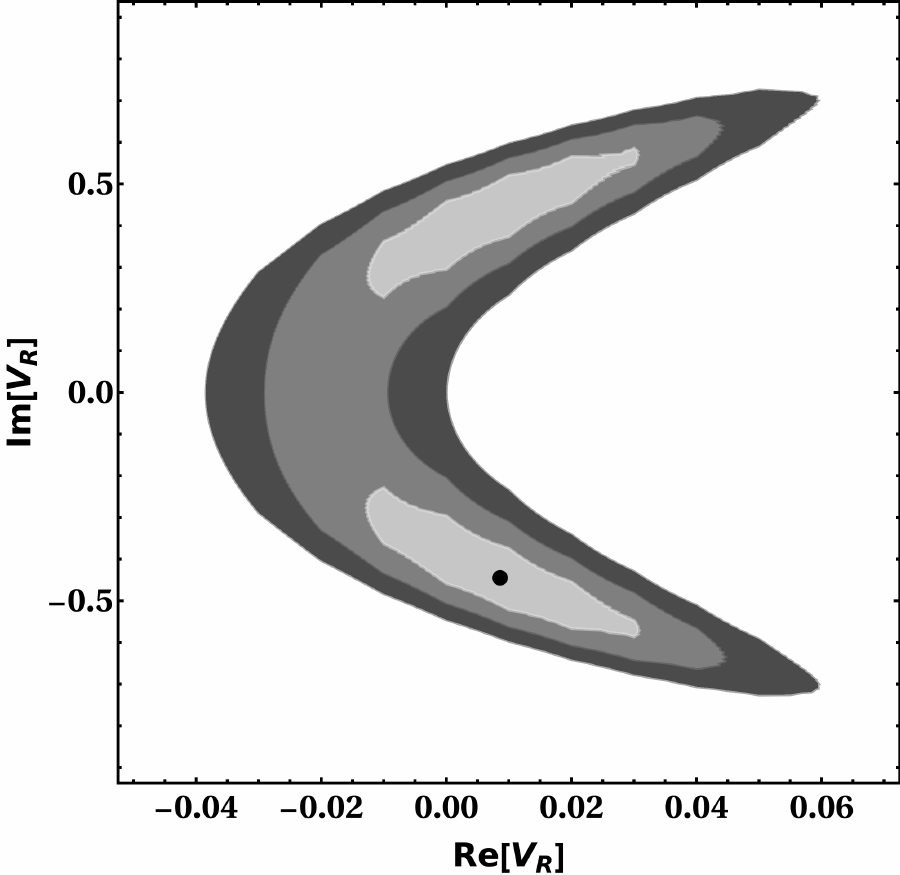}\\
        \includegraphics[width=0.38\linewidth]{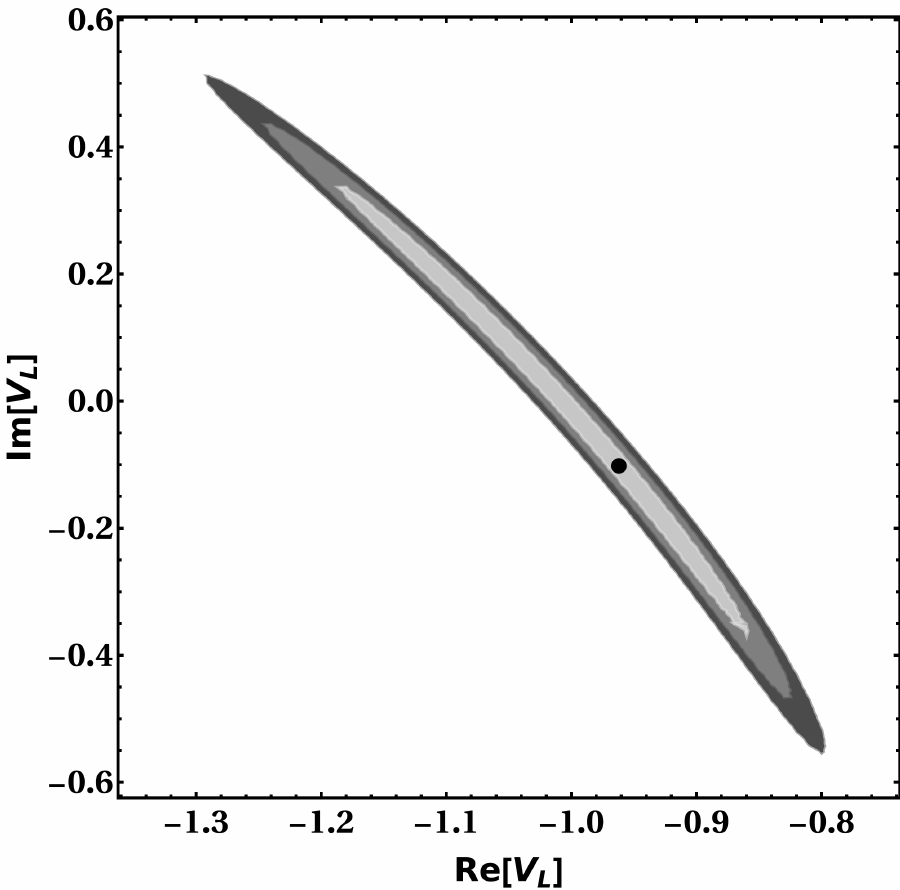}\quad
         \includegraphics[width=0.38\linewidth]{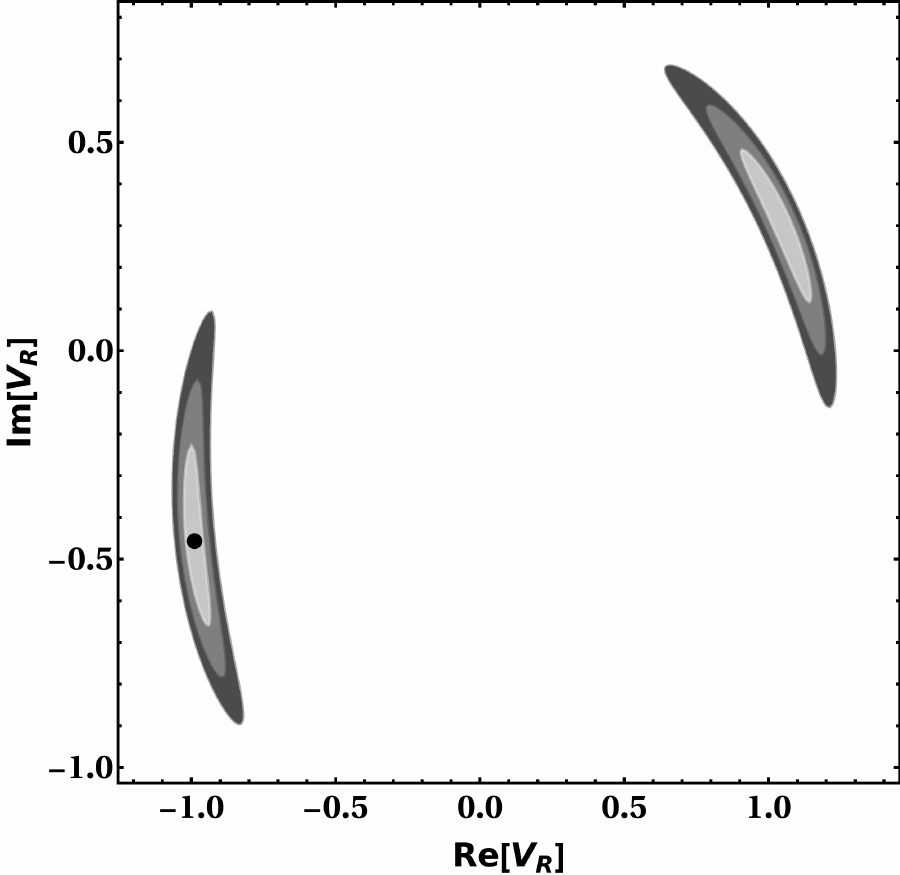}\\
    \caption{ $\chi^2$ analysis plots for new physics scenarios in Case \textbf{A}, including axial(vector) coefficients, depicting the fit quality and confidence regions.}
    \label{fig:CA-contours}
\end{figure}
%%%%%%%%%%%%%%

\begin{figure}[htp]
    \centering
    \includegraphics[width=0.42\linewidth]{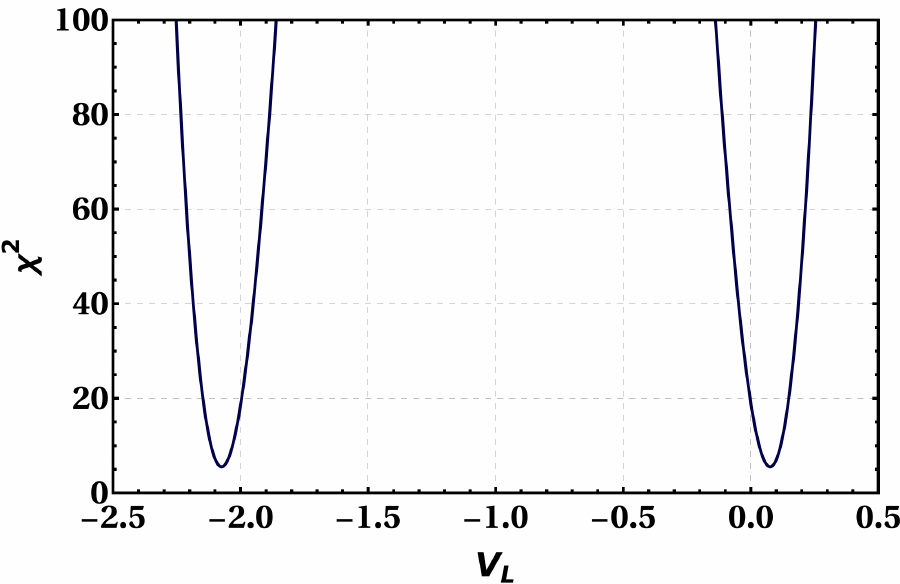} \quad
    \includegraphics[width=0.42\linewidth]{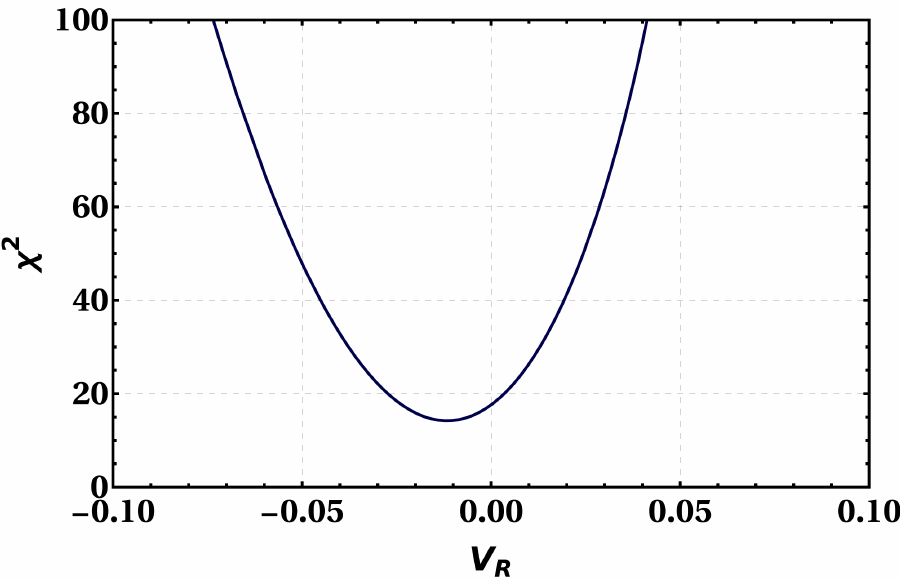}\\
    \includegraphics[width=0.38\linewidth]{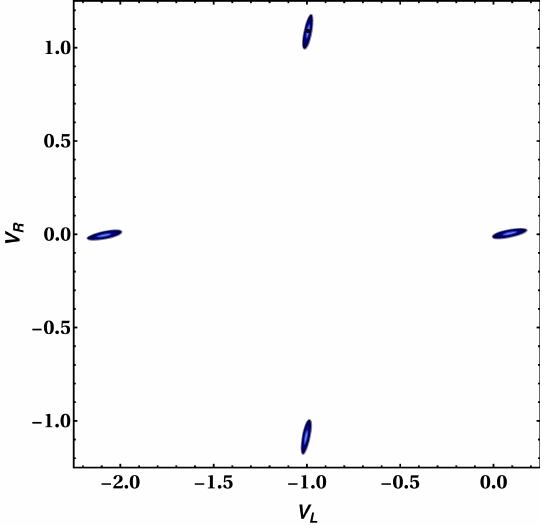}\quad
     \includegraphics[width=0.38\linewidth]{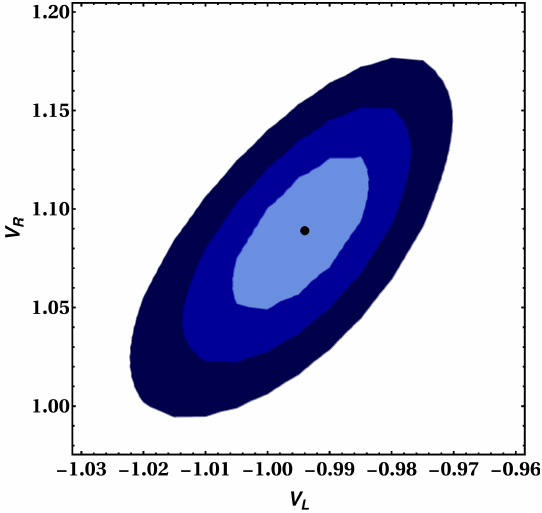}\\
      \includegraphics[width=0.38\linewidth]{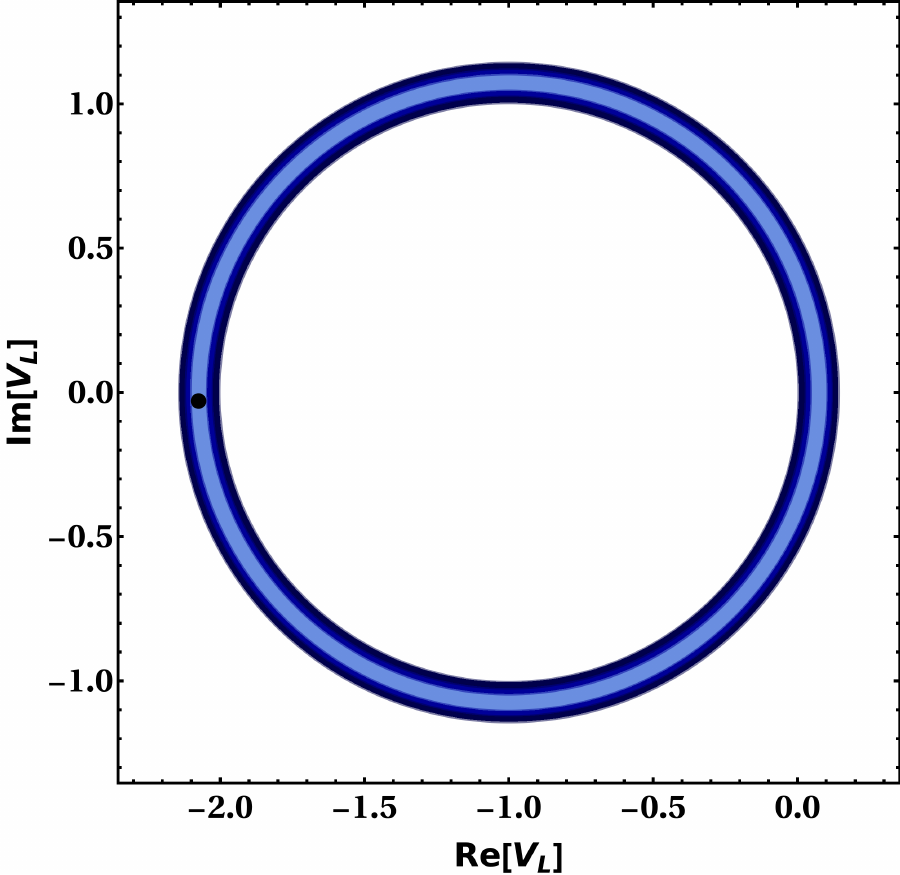}\quad
       \includegraphics[width=0.38\linewidth]{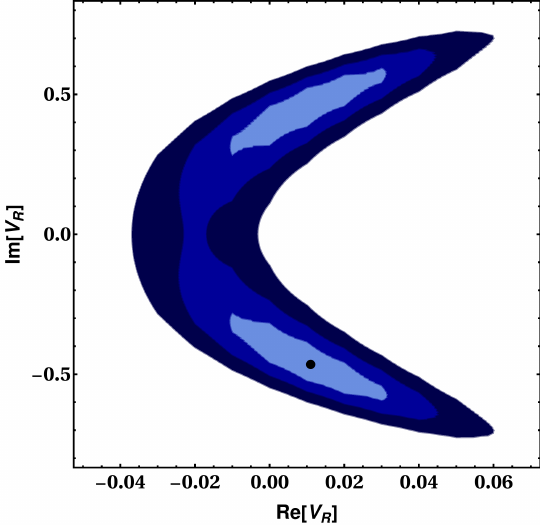}\\
        \includegraphics[width=0.38\linewidth]{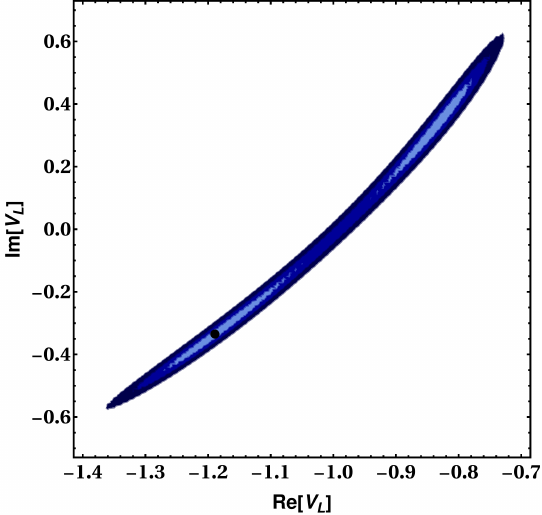}\quad
         \includegraphics[width=0.38\linewidth]{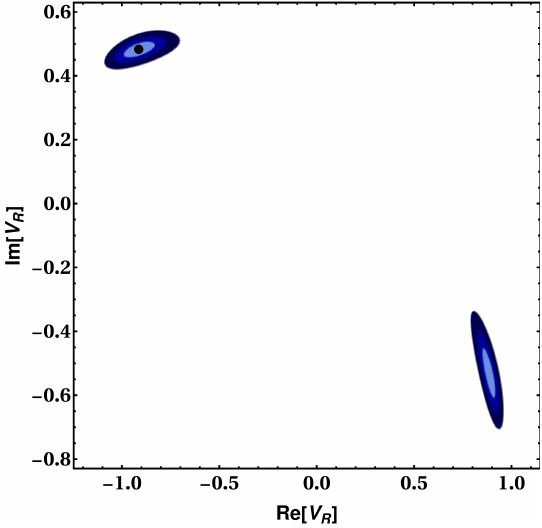}\\
    \caption{Same as Fig. \ref{fig:CA-contours} for the Case \textbf{B}.}
    \label{fig:CB-contours}
\end{figure}
%=============================================================================
After performing the global fit analysis for both Case \textbf{A} and Case \textbf{B}, we report the best-fit values along with their pull, defined as $\sqrt{\chi_{\text{SM}}^2 - \chi^2_{\text{best-fit}}}$, and the p-values (in $\%$) presented in Table \ref{Tab:bestfit}.  The pull value quantifies the improvement of the fit when NP contributions are included, compared to the fit with the SM alone. The $p$-value is defined as the probability, under the assumption that the null hypothesis is true, of obtaining a test statistic at least as extreme as the one observed. In a $\chi^2$ fit, it quantifies the probability that the observed (or a larger) value of $\chi^2_{\text{min}}$ could arise from statistical fluctuations. If all observables were ``clean,'' i.e., with negligible theoretical uncertainties, the $\chi^2_{\text{min}}/\text{d.o.f.}$ distribution would be approximately Gaussian with a central value of 1, corresponding to a $p$-value of 0.5. In general, a fit with $p < 0.05$ (outside the 95\% confidence level region) is considered to provide an unacceptable description of the data. The $1\sigma$, $2\sigma$, and $3\sigma$ confidence level contour plots of the new parameters for all scenarios in Case~\textbf{A} and Case~\textbf{B} are shown in Figs.~\ref{fig:CA-contours} and~\ref{fig:CB-contours}, respectively. These contours are obtained by evaluating the change in $\chi^2$ with respect to the global minimum, where the standard values $\Delta\chi^2 = 2.30$, $6.18$, and $11.83$ correspond to the $1\sigma$, $2\sigma$, and $3\sigma$ regions, respectively, for two degrees of freedom. The best-fit point is indicated by a black dot in each figure and corresponds directly to the central values reported in Table~\ref{Tab:bestfit}. Thus, the contours illustrate the allowed parameter space around the best-fit point, while Table~\ref{Tab:bestfit} provides the corresponding numerical best-fit values. We observed that the percentage of the p-value has increased for the combined fit of the $b \to (u, c) \tau \bar{\nu}_\tau$ observables to the new parameters. The possible combinations of 2D coefficients yield the best p-values, indicating the best fit to the experimental data. Thus, we are going to present the impact of these new physics scenarios of (axial)vector Wilson coefficients on the $B_s \to D_s^{(*)} \tau \bar \nu_\tau$ decay modes for Case \textbf{B}.
%%%%%%%%%%%%%%%%%%
\begin{table}[htb]
%\flushleft \flushleft \small
\centering
\begin{tabular}{|l | l | l | l | l |}
\toprule[1.2pt] 
~\textbf{Scenario}~ & ~\textbf{Coefficient}~ & ~\textbf{Best-fit value}  ~& ~\textbf{Pull} ~&~  \textbf{p-value} ($\%$) ~\\
\hline
\multicolumn{5}{|c|}{\textbf{Case A}} \\
\hline
 %\midrule 
1DS - I & $Re[V_L]$ &  $-2.07$   & $3.40$ & $62.41$\\ 
1DS - II & $Re[V_R]$ &  $-0.01$   & $1.75$ & $3.55$ \\ 
2DS - I & $(Re[V_L],Re[V_R])$ & $(-2.09, -0.005)$  & $3.46$ & $55.36$ \\ 
2DS - II & $(Re[V_L], Im[V_L])$&  $(0.03, -0.29)$ &  $3.40$  & $47.86$ \\ 
2DS - III & $(Re[V_R], Im[V_R])$ & $(0.009, -0.45)$ & $3.44$  & $52.82$  \\ 
4DS - I & $(Re[V_L], Im[V_L], Re[V_R], Im[V_R])$ & $(-0.96, -0.1, -0.99, -0.46)$ & $3.44$  & $20.39$   \\ 
\bottomrule[1.2pt] 
\multicolumn{5}{|c|}{\textbf{Case B}} \\
\hline
1DS - I & $Re[V_L]$ &  $-2.08$   & $3.65$ & $69.50$\\ 
1DS - II & $Re[V_R]$ &  $-0.01$   & $1.85$ & $7.58$\\ 
2DS - I & $(Re[V_L],Re[V_R])$ &  $(-0.99, 1.09)$ & $3.72$  & $80.53$  \\ 
2DS - II & $(Re[V_L], Im[V_L])$& $(-2.07, -0.03)$ & $3.65$  & $73.81$ \\ 
2DS - III & $(Re[V_R], Im[V_R])$ & $(0.01, -0.47)$ &  $3.72$   & $80.25$\\ 
4DS - I & $(Re[V_L], Im[V_L], Re[V_R], Im[V_R])$ & $(-1.89, -0.34, -0.91, 0.48)$ & $3.72$ & $57.85$  \\
%& $Re[V_R], Im[V_R])$ & $-0.35[\substack{-0.34\\-0.36}], -0.632 [\substack{-0.622\\ -0.642}])$, & $3.18$ & $50.55$  \\
\bottomrule[1.2pt] 
\end{tabular}
\caption{ Best-fit, pull and p-value($\%$) of various new physics scenarios.}\label{Tab:bestfit}
\end{table}
%%%%%%%%%%%%%%%%%%%%%%%%%%%%%%%%%%%

%============================================
\section{$B_s \to D_s^{(*)} \tau \bar \nu_\tau$ decay modes}
%===========================================

%=================================================
\subsection{Mathematical Expressions for Key Observables}
%================================================
The branching ratio of the $\bar{B_s} \to D_s \tau \bar{\nu}_\tau$ process as a function of $q^2$ in the presence of new (axial)vector coefficients is given by \cite{Sakaki:2013bfa}
\bea 
\frac{d{\rm BR}(\bar B_s \to  D_s \tau \bar \nu_\tau)}{dq^2} &=& \tau_B {G_F^2 |V_{cb}|^2 \over 192\pi^3 M_B^3} q^2 \sqrt{\lambda_{D_s}(q^2)} \left( 1 - {m_\tau^2 \over q^2} \right)^2  \nn \\   && \times \Big | 1 + V_L + V_R \Big |^2 \left[ \left( 1 + {m_\tau^2 \over 2q^2} \right) H_{0}^{2} + {3 \over 2}{m_\tau^2 \over q^2}  H_{t}^{2} \right]\,,
\eea
where $\lambda_{D_s} =\lambda (M_B^2, M_{D_s}^2, q^2), ~ {\rm with}~\lambda(a,b,c)=a^2+b^2+c^2-2(ab+bc+ca)$
 and  $H_{0,t}$'s are the heicity amplitudes given in the appendix \ref{A}.

 The branching ratios of $\bar B \to D_s^* \tau \bar \nu_l$  with respect to $q^2$ is given by \cite{Sakaki:2013bfa}
\bea
 {d{\rm BR}(\bar B_s \to  D_s^* \tau \bar \nu_\tau) \over dq^2} &=& \tau_B{G_F^2 |V_{cb}|^2 \over 192\pi^3 M_B^3} q^2 \sqrt{\lambda_{D_s^*} (q^2)} \left( 1 - {m_\tau^2 \over q^2} \right)^2 \times \nn  \\ && \bigg \{  \left( \left|1 + V_L \right|^2 + \left| V_R\right|^2 \right)  \left[ \left( 1 + {m_\tau ^2 \over 2q^2} \right) \left( H_{V, +}^2 + H_{V,-}^2 + H_{V,0}^2 \right) + {3 \over 2}{m_\tau^2 \over q^2} \, H_{V,t}^2 \right]  \nn \\ && - 2{\rm Re}\left[\left(1+ V_L \right) V_R^* \right] \left[ \left( 1 + {m_\tau^2 \over 2q^2} \right) \left( H_{V,0}^2 + 2 H_{V,+} H_{V,-} \right) + {3 \over 2}{m_\tau^2 \over q^2} \, H_{V,t}^2 \right] \bigg \},\,\, ~~~~
\eea
where  $\lambda_{D_s^*}= \lambda (M_B^2, M_{D_s^*}^2, q^2)$ and  $H_{V,\lambda}$'s ($\lambda=+,-,0$) are the helicity amplitudes presented in appendix \ref{A}. 

In addition to the branching ratios, we also examine the following observables \cite{Sakaki:2013bfa} to investigate the structure of new physics.
\begin{itemize}
\renewcommand{\labelitemi}{--}
\item Lepton non-universality:
\bea
&&R_{D_s^{(*)}}=\frac{{{\rm BR}}( B_s \to D_s^{(*)} \tau \bar \nu_\tau)}{{{\rm BR}}(B_s \to D_s^{(*)} l \bar \nu_l)}, ~~~~l=e, \mu.
\eea 

\item $\tau$ forward-backward asymmetry:
\bea
A_{\rm FB}^{D_s^{(*)}} = { \int_0^1 {d\Gamma \over d\cos\theta}d\cos\theta-\int^0_{-1}{d\Gamma \over d\cos\theta}d\cos\theta \over \int_{-1}^1 {d\Gamma \over d\cos\theta}d\cos\theta }\,.
\eea
\item $\tau$ forward and backward fractions:
\bea
\chi_{1,2}^{D_s^{(*)}}=\frac{1}{2}R_{D_s^{(*)}}\left(1\pm A_{FB}^{D_s^{(*)}}\right)\,.
\eea
\item $\tau$ polarization asymmetry:
\bea
P_\tau^{D_s^{(*)}} (q^2) = \frac{ d\Gamma (\lambda_\tau = 1/2)/dq^2 - d\Gamma (\lambda_\tau = -1/2)/dq^2}{d\Gamma (\lambda_\tau = 1/2)/dq^2 + d\Gamma (\lambda_\tau = -1/2)/dq^2}\,.
\eea
\item $\tau$ spin $1/2$ and $-1/2$ fractions:
\bea
\chi_{3,4}^{D_s^{(*)}}=\frac{1}{2}R_{D_s^{(*)}} \left(1\pm P_\tau^{D_s^{(*)}}\right)\,.
\eea
\item $D_s^*$ polarization asymmetry:
\bea
F_{L, T}^{D_s^*}(q^2) = \frac{d\Gamma_{L, T} \left(B_s \to D_s^* \tau \bar{\nu} \right)/ dq^2}{d\Gamma \left(B_s \to D_s^* \tau \bar{\nu} \right) / dq^2}\,.
\eea
\item $D_s^*$ longitudinal and transverse polarization fractions:
\bea
\chi_{5,6}^{D_s^{*}}=R_{D_s^{*}}F_{L,T}^{D_s^*}\,.
\eea
\end{itemize}

%==============================================
\subsection{Results and Discussion}
%====================================================

\begin{figure}[htb]
    \centering
    \includegraphics[width=0.48\linewidth]{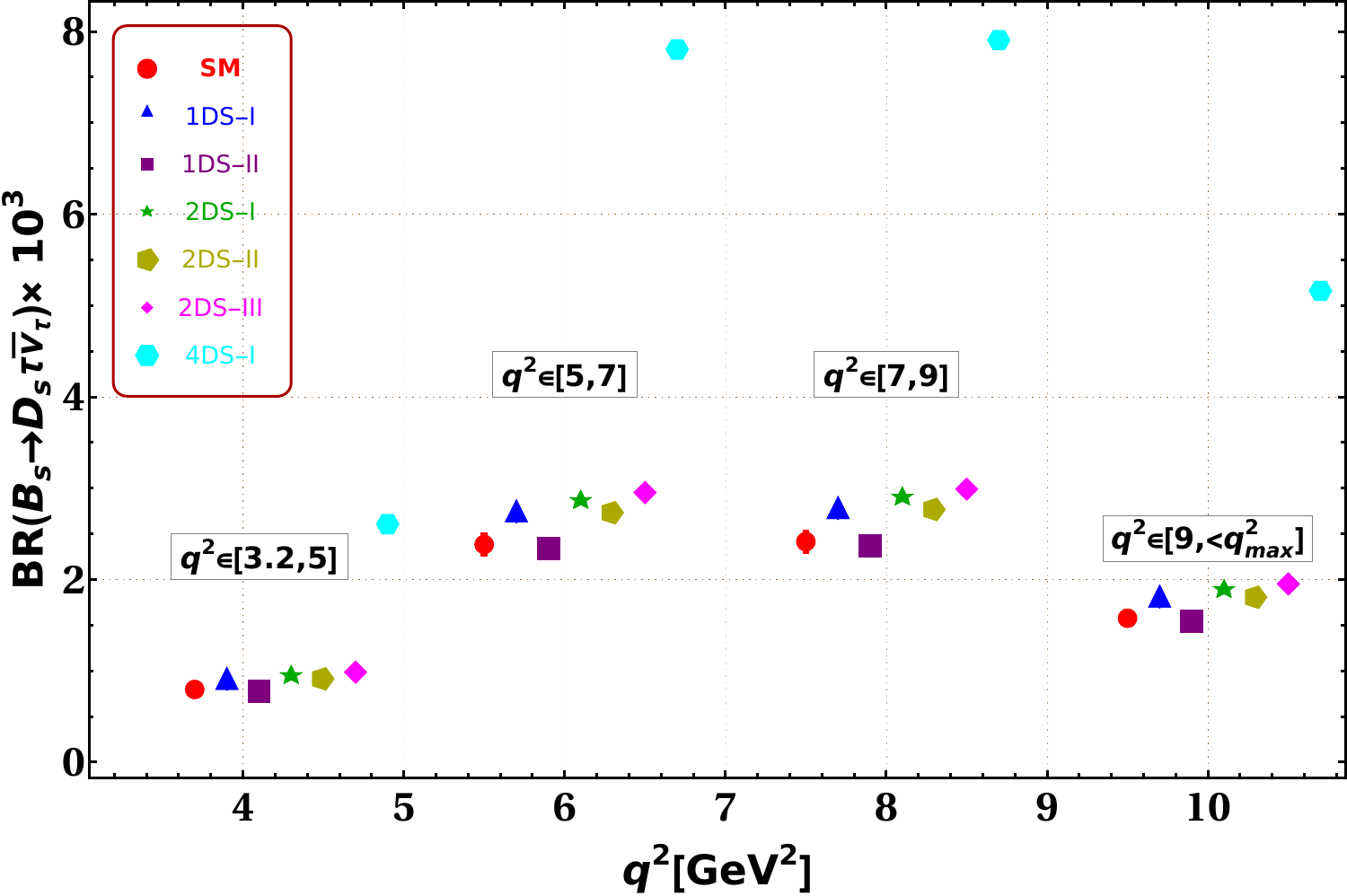} \quad
    \includegraphics[width=0.48\linewidth]{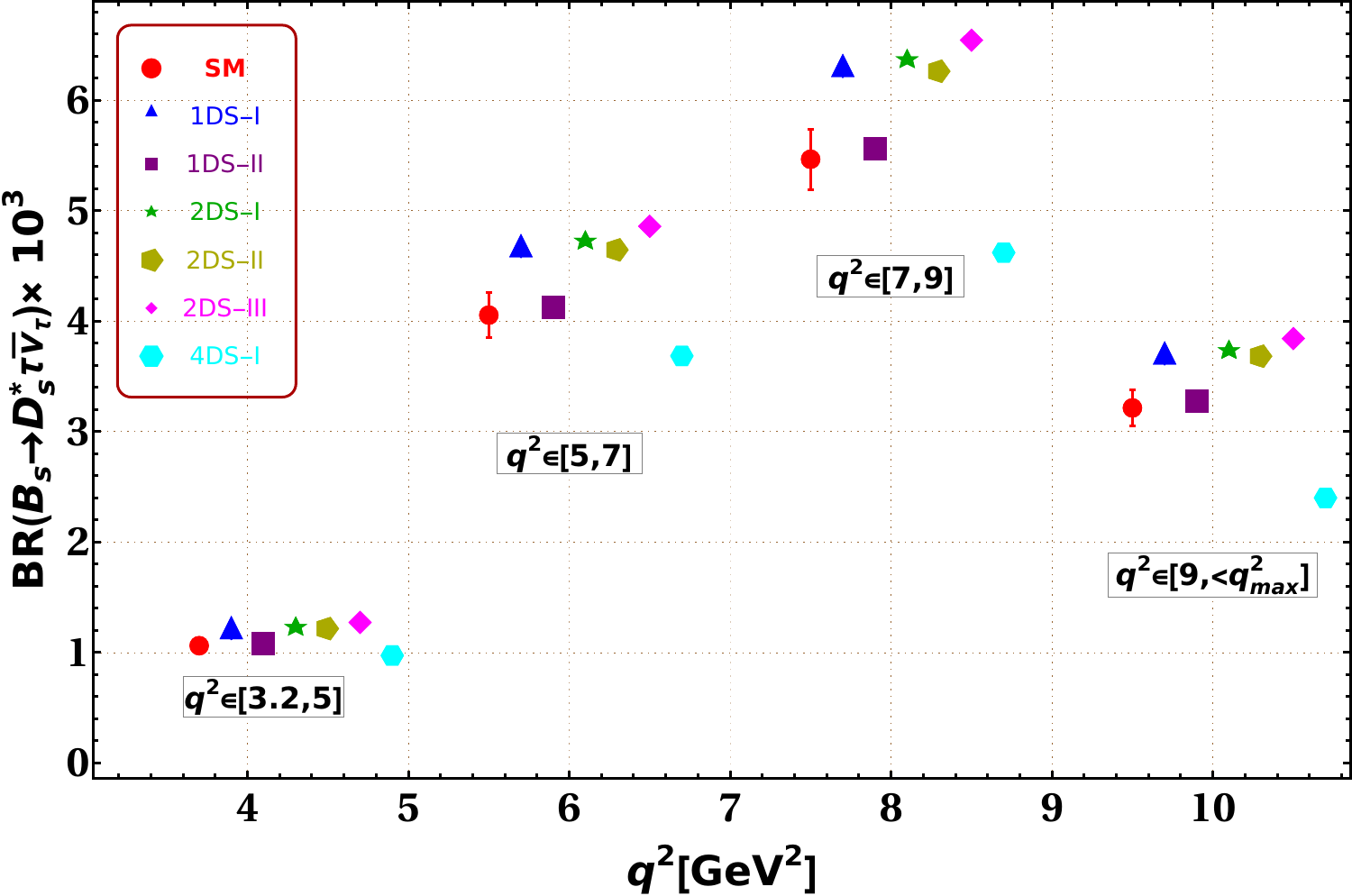}
    \caption{The $q^2$ binwise predictions for the branching ratios of $B_s \to D_s \tau \bar{\nu}_\tau$ (left panel) and $B_s \to D_s^* \tau \bar{\nu}_\tau$ (right panel) decay processes are shown for all new physics scenarios. The red circle represents the SM central value prediction, and the red error lines indicate the $1\sigma$ theoretical uncertainties. Predictions from the various new physics scenarios are presented in different colors.}
    \label{fig:BR}
\end{figure}

\begin{figure}[htb]
    \centering
    \includegraphics[width=0.48\linewidth]{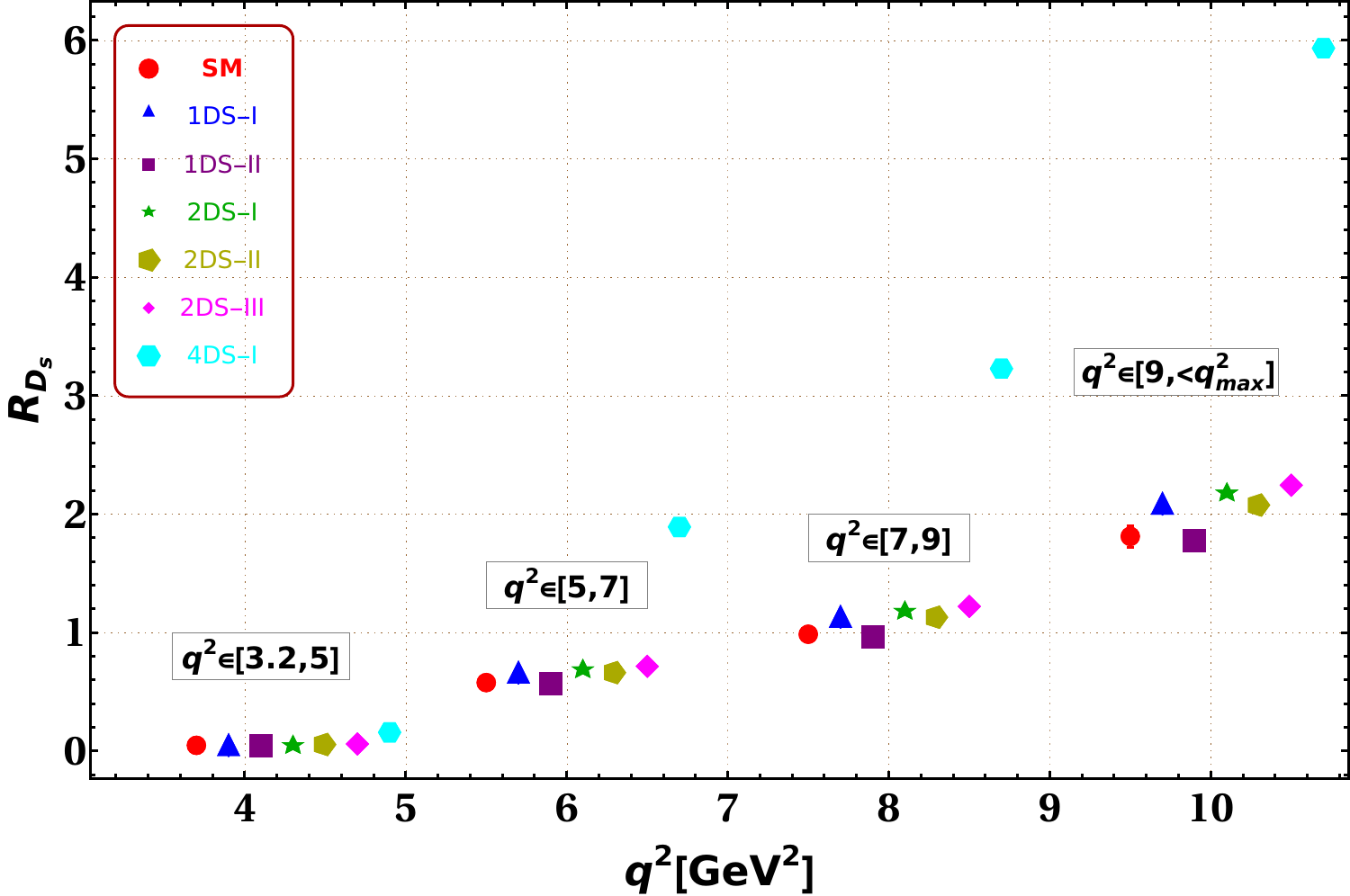} 
    \includegraphics[width=0.48\linewidth]{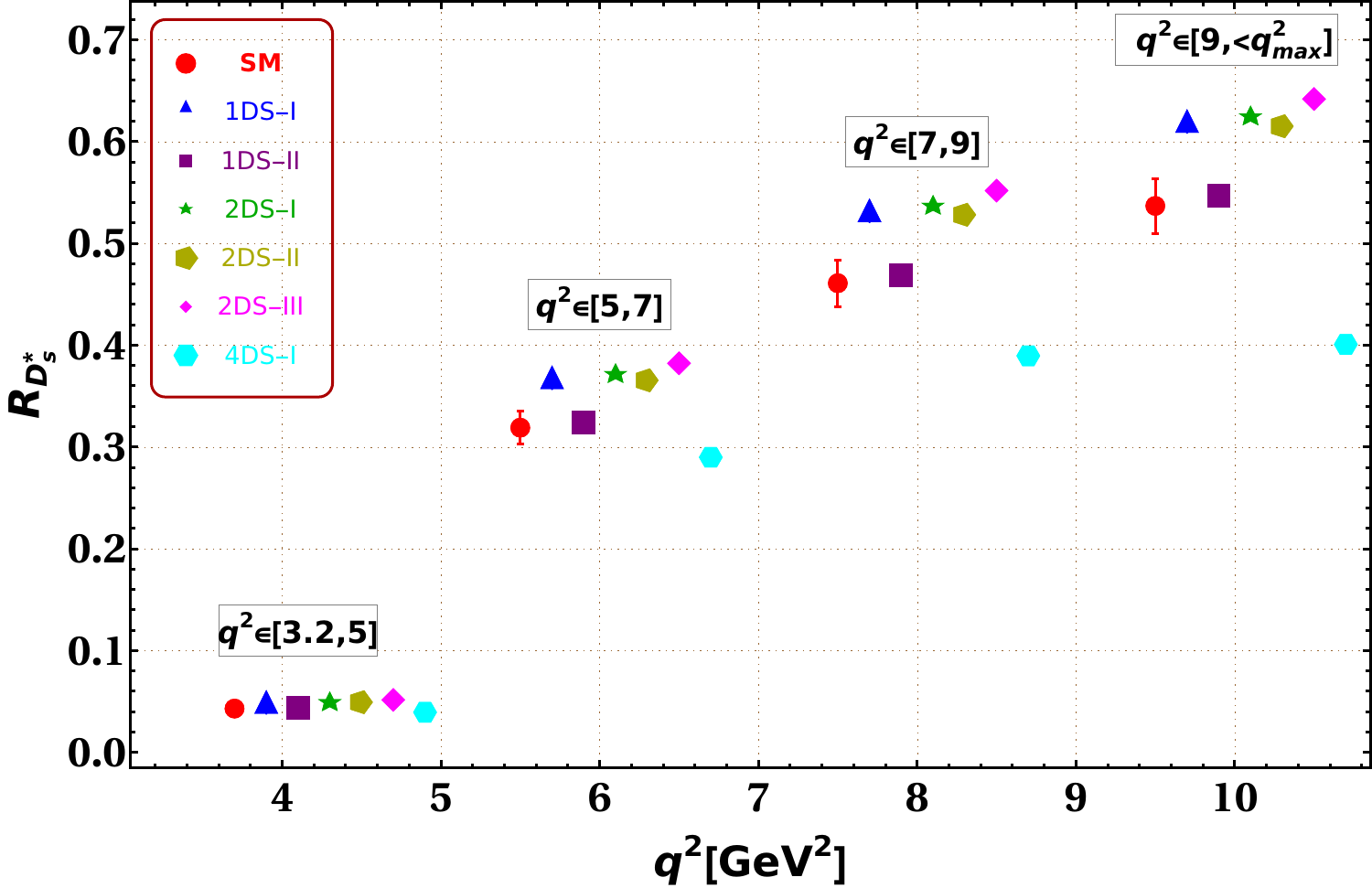}
    \caption{Same as Fig. \ref{fig:BR} for $R_{D_s}$ (left panel) and $R_{D_s^*}$ (right panel).}
    \label{fig:LNU}
\end{figure}

\begin{figure}[htb]
    \centering
    \includegraphics[width=0.48\linewidth]{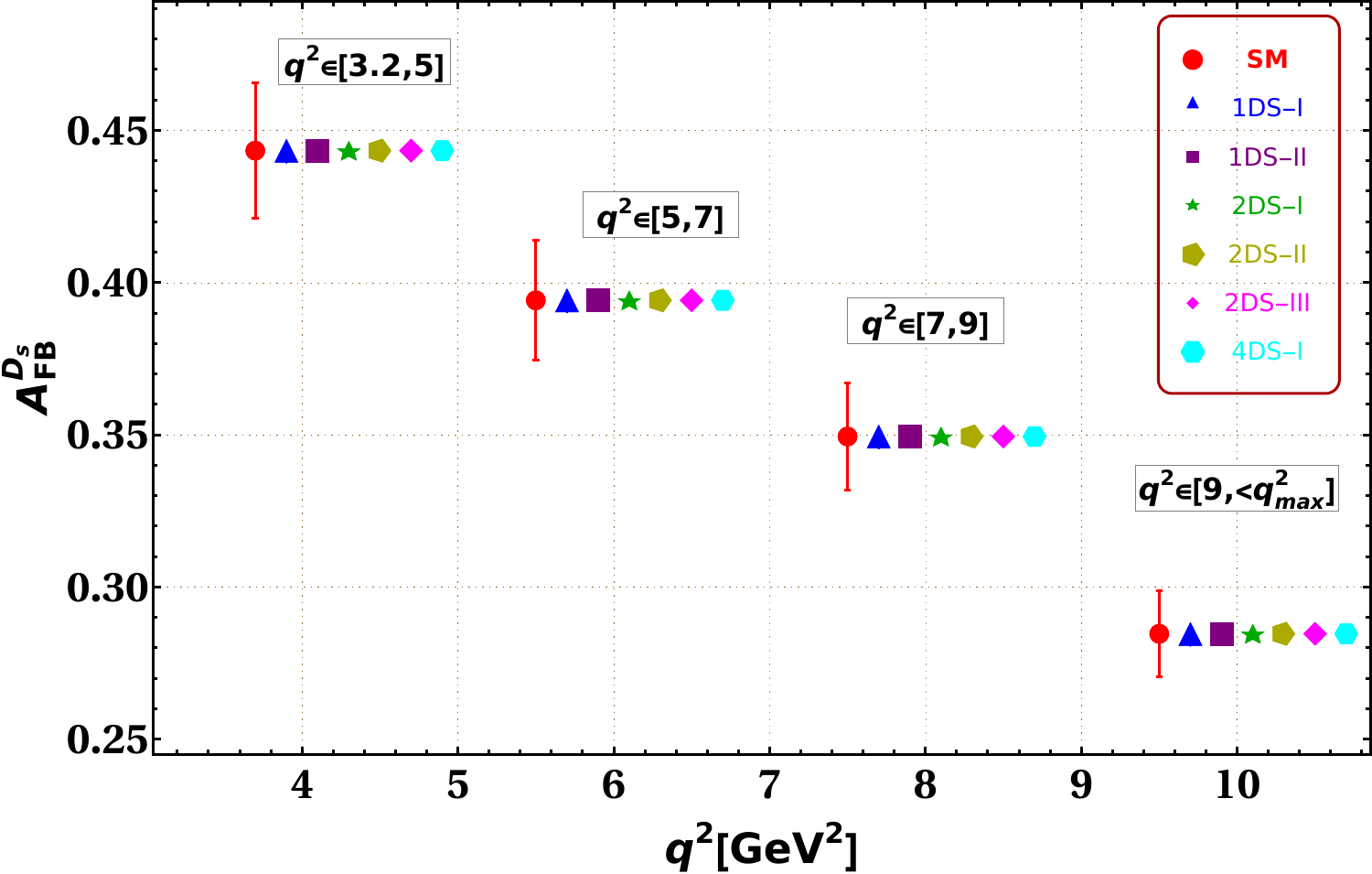} \quad
     \includegraphics[width=0.48\linewidth]{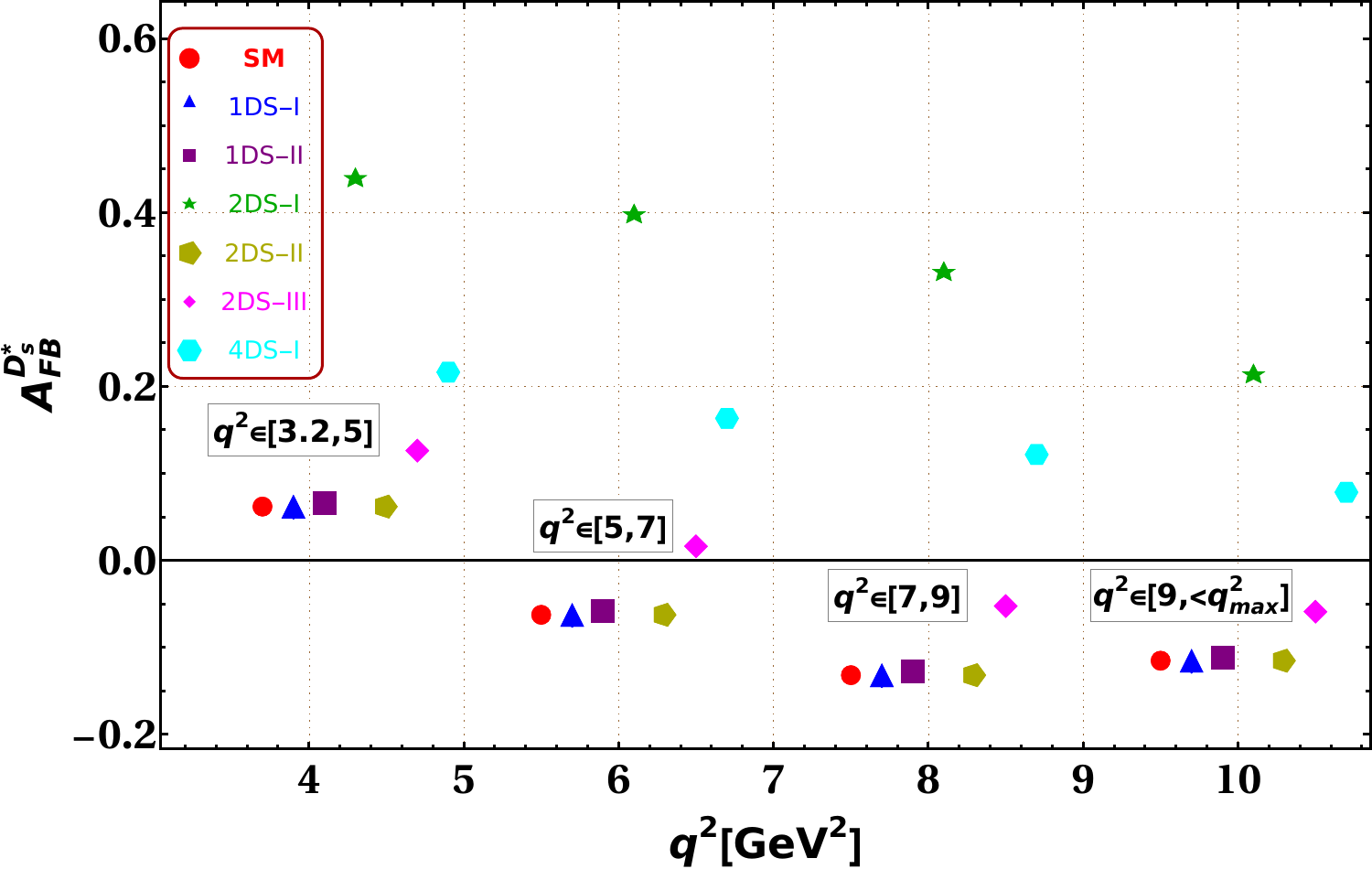}
    \includegraphics[width=0.48\linewidth]{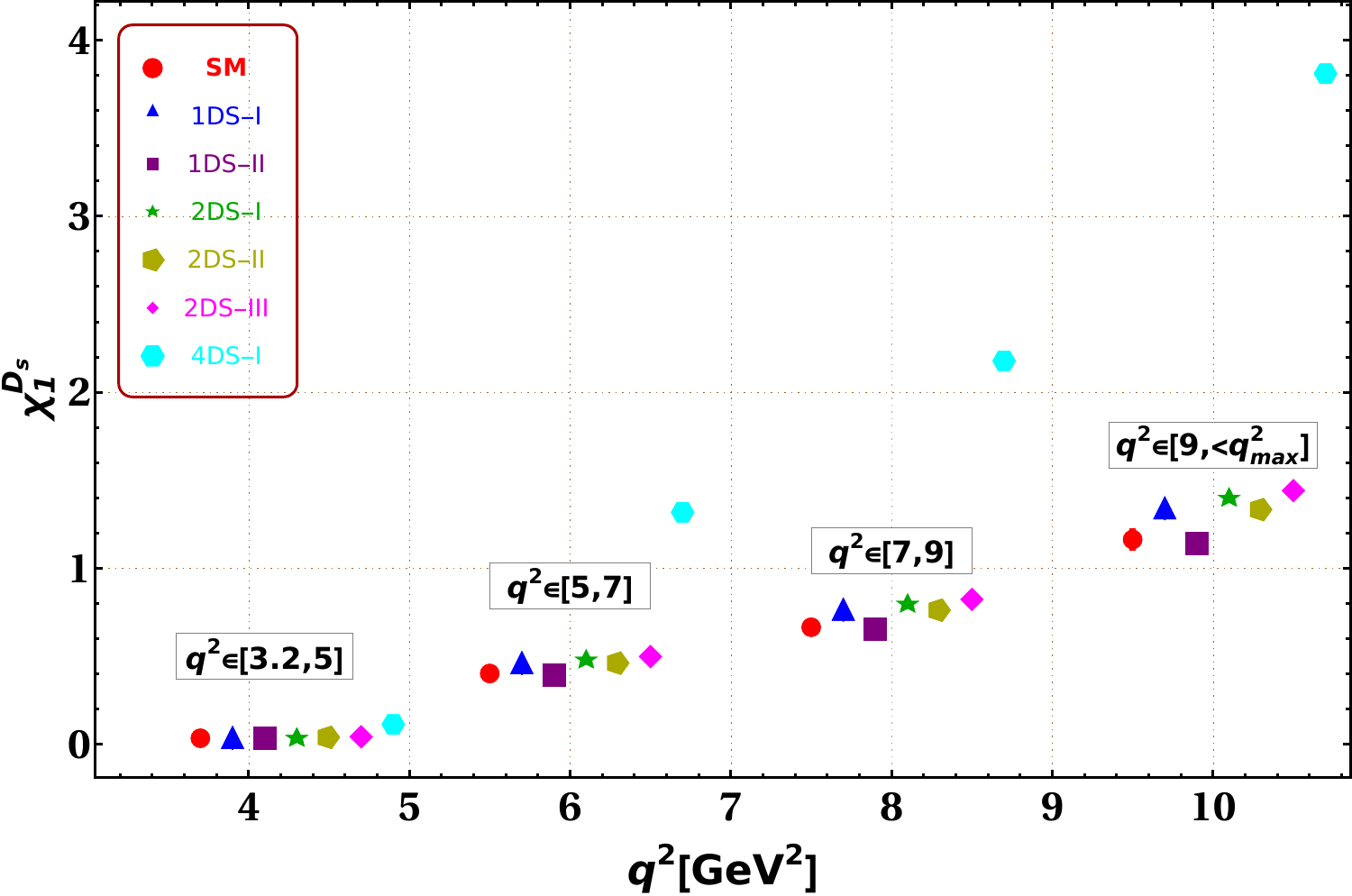} \quad
    \includegraphics[width=0.48\linewidth]{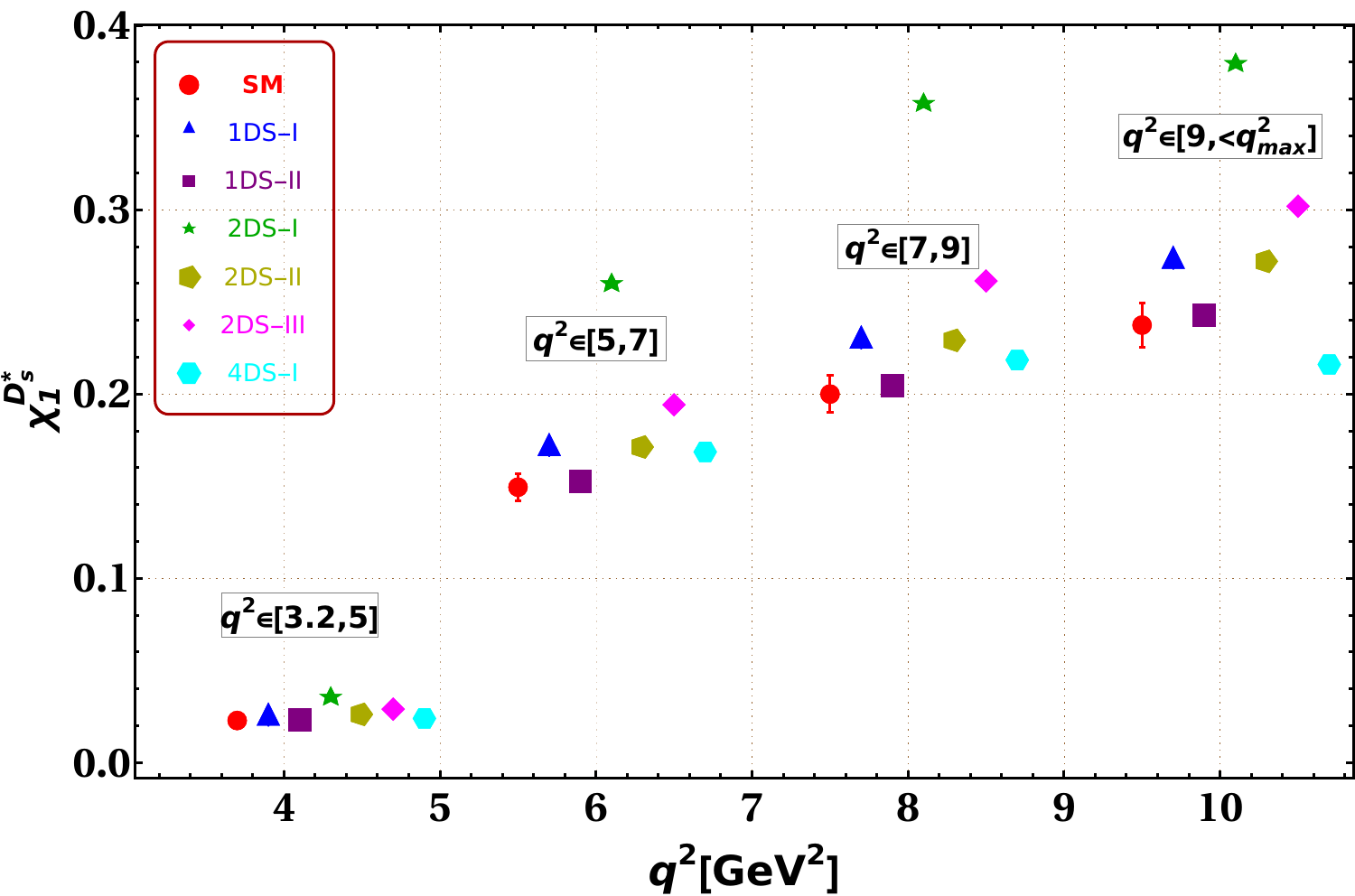}
     \includegraphics[width=0.48\linewidth]{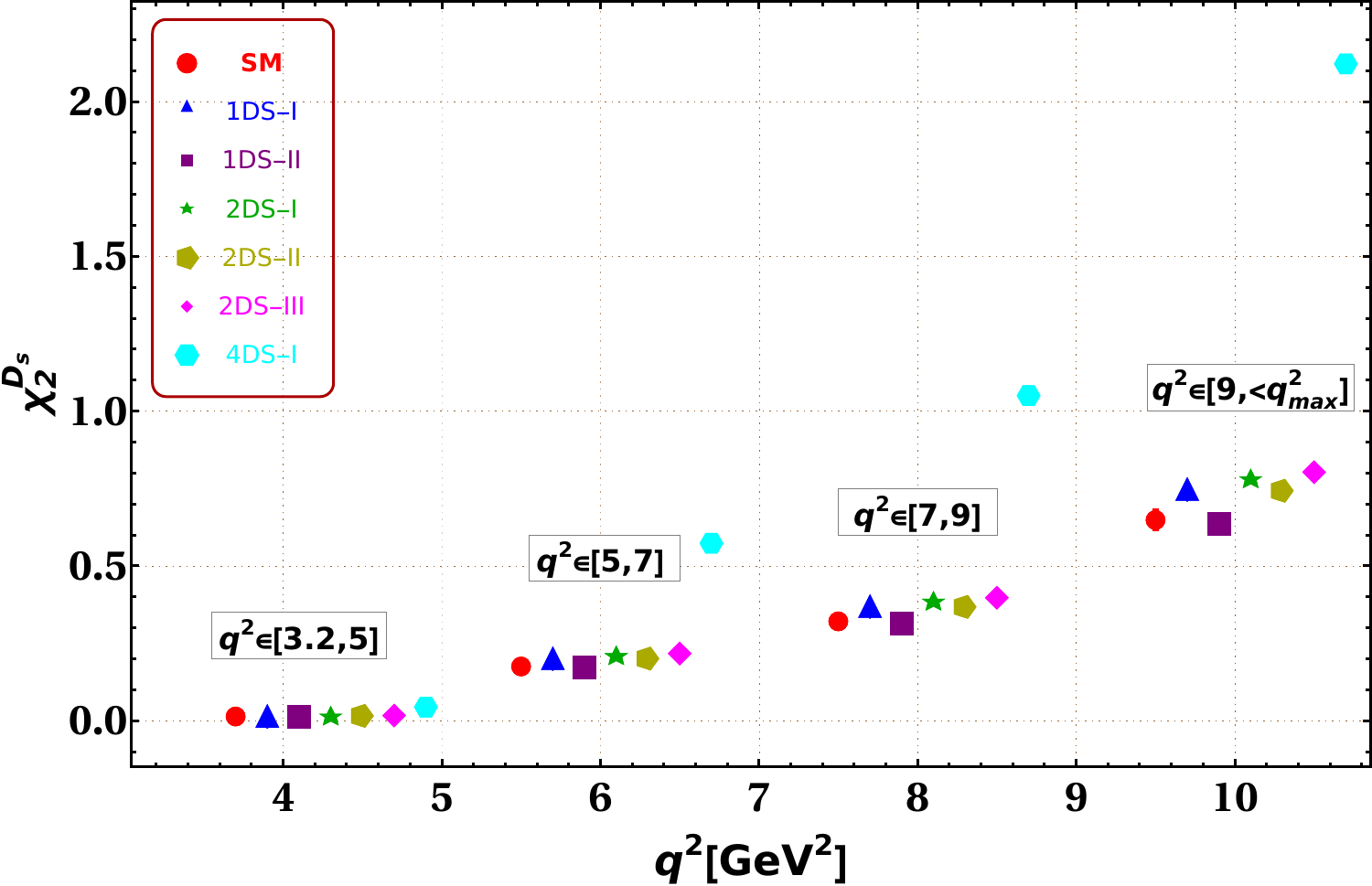} \quad
      \includegraphics[width=0.48\linewidth]{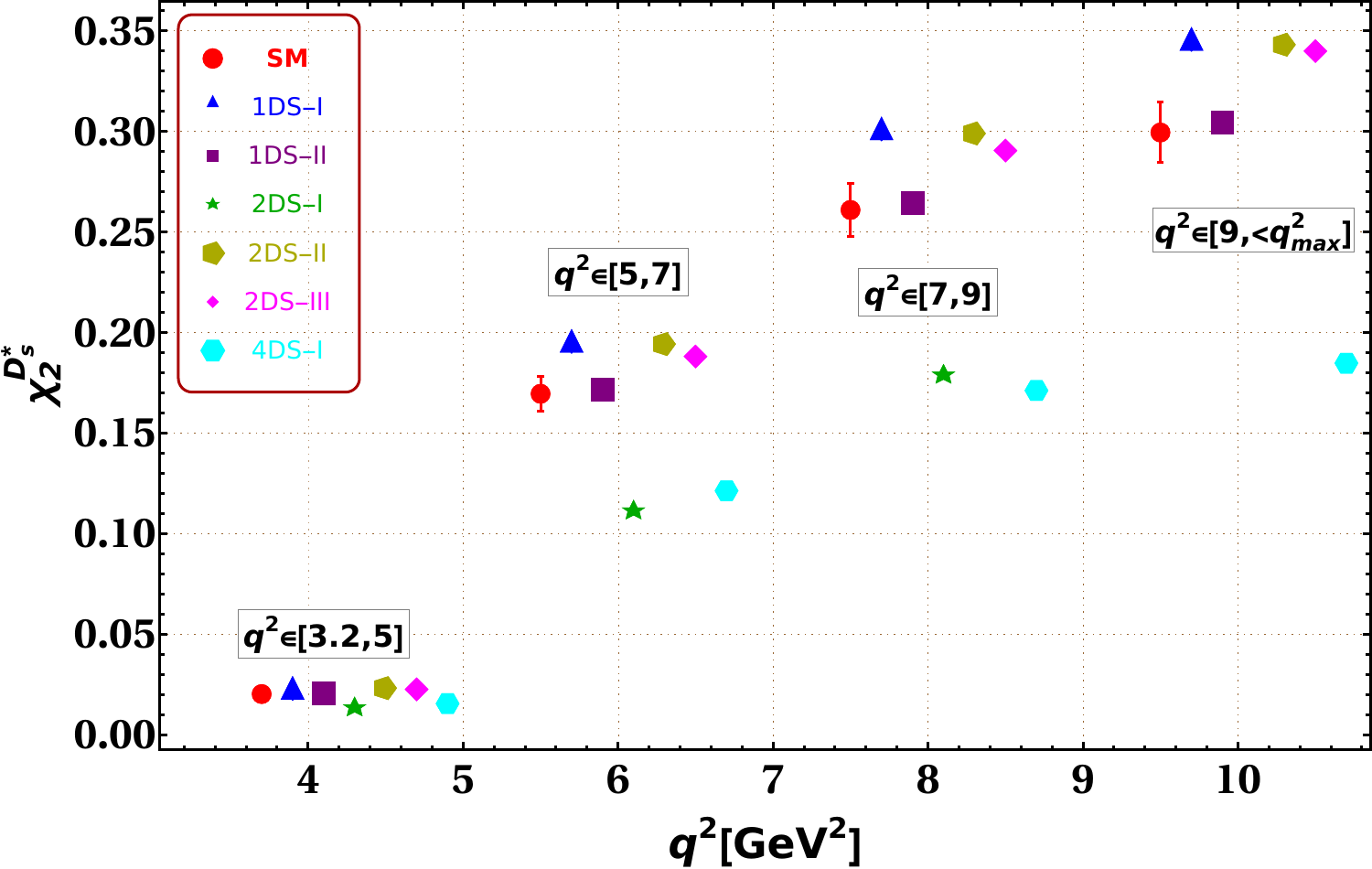} 
    \caption{Same as Fig. \ref{fig:BR} for $A_{FB}^{D_s}$ (top left panel), $A_{FB}^{D_s^*}$ (top right panel), $\chi_1^{D_s}$ (middle left panel), $\chi_1^{D_s^*}$ (middle right panel), $\chi_2^{D_s}$ (bottom left panel) and  $\chi_2^{D_s^*}$ (bottom right panel).}
    \label{fig:AFB}
\end{figure}

\begin{figure}[htb]
    \centering
    \includegraphics[width=0.48\linewidth]{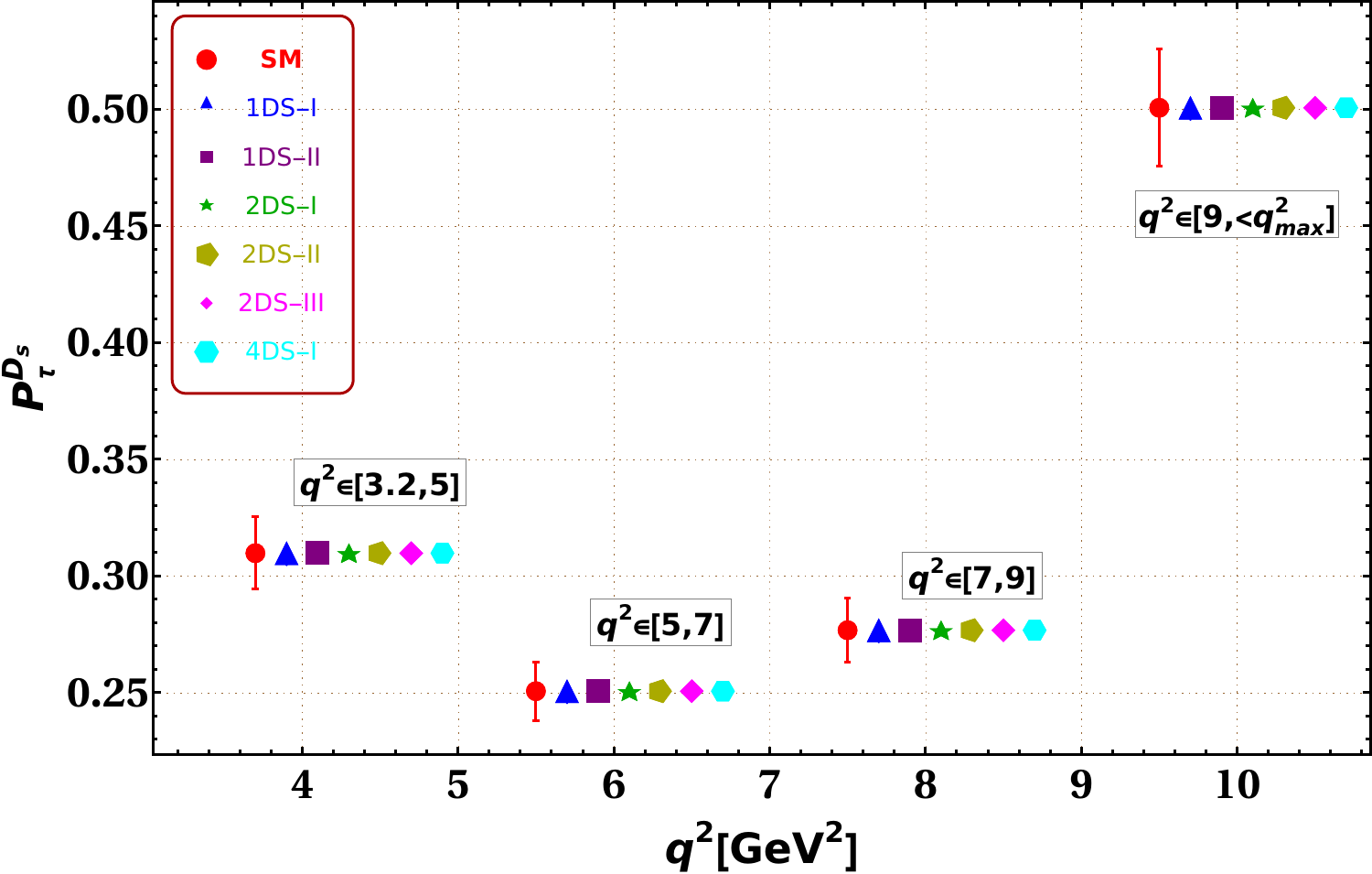} \quad
     \includegraphics[width=0.48\linewidth]{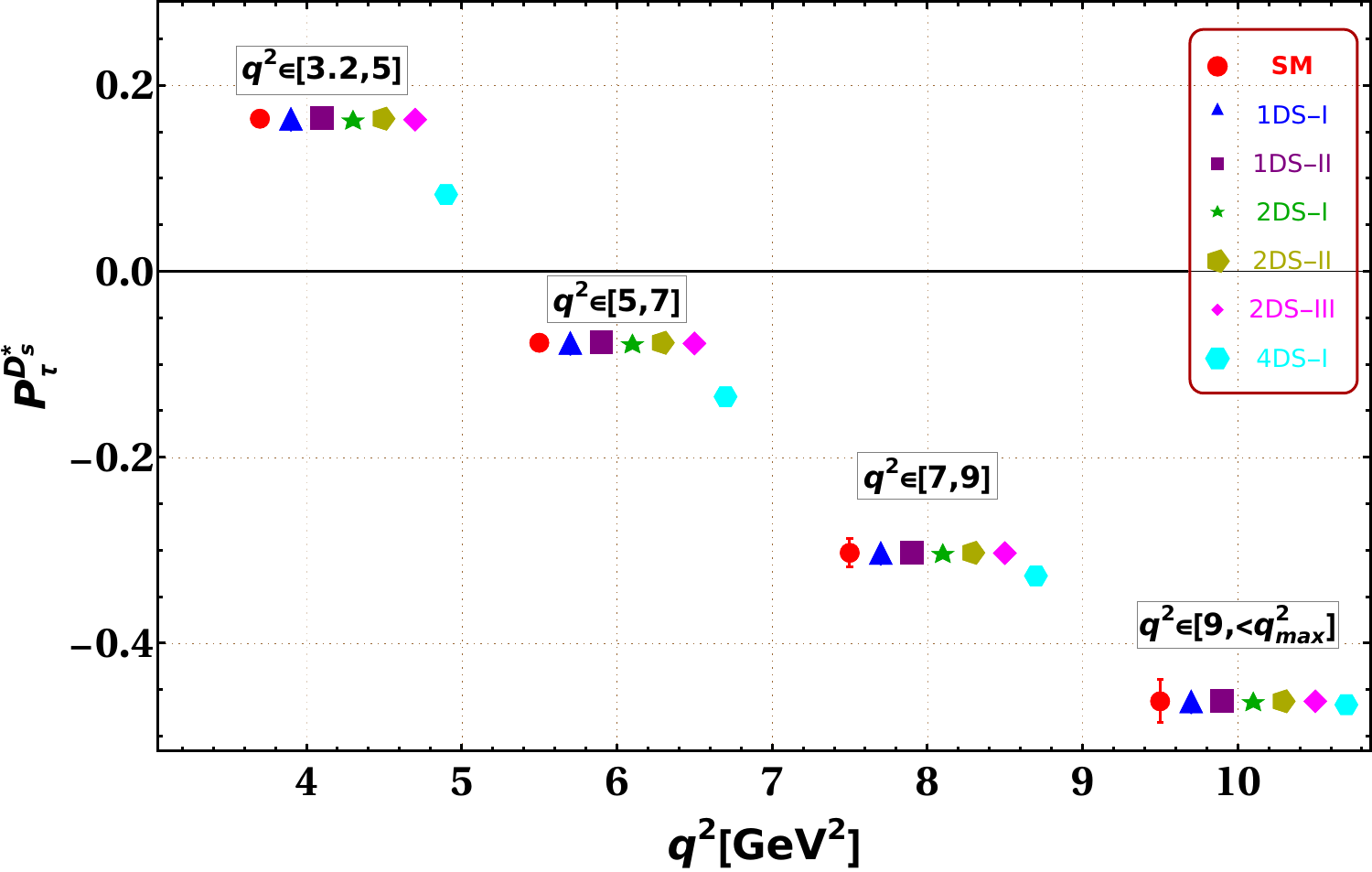} 
    \includegraphics[width=0.48\linewidth]{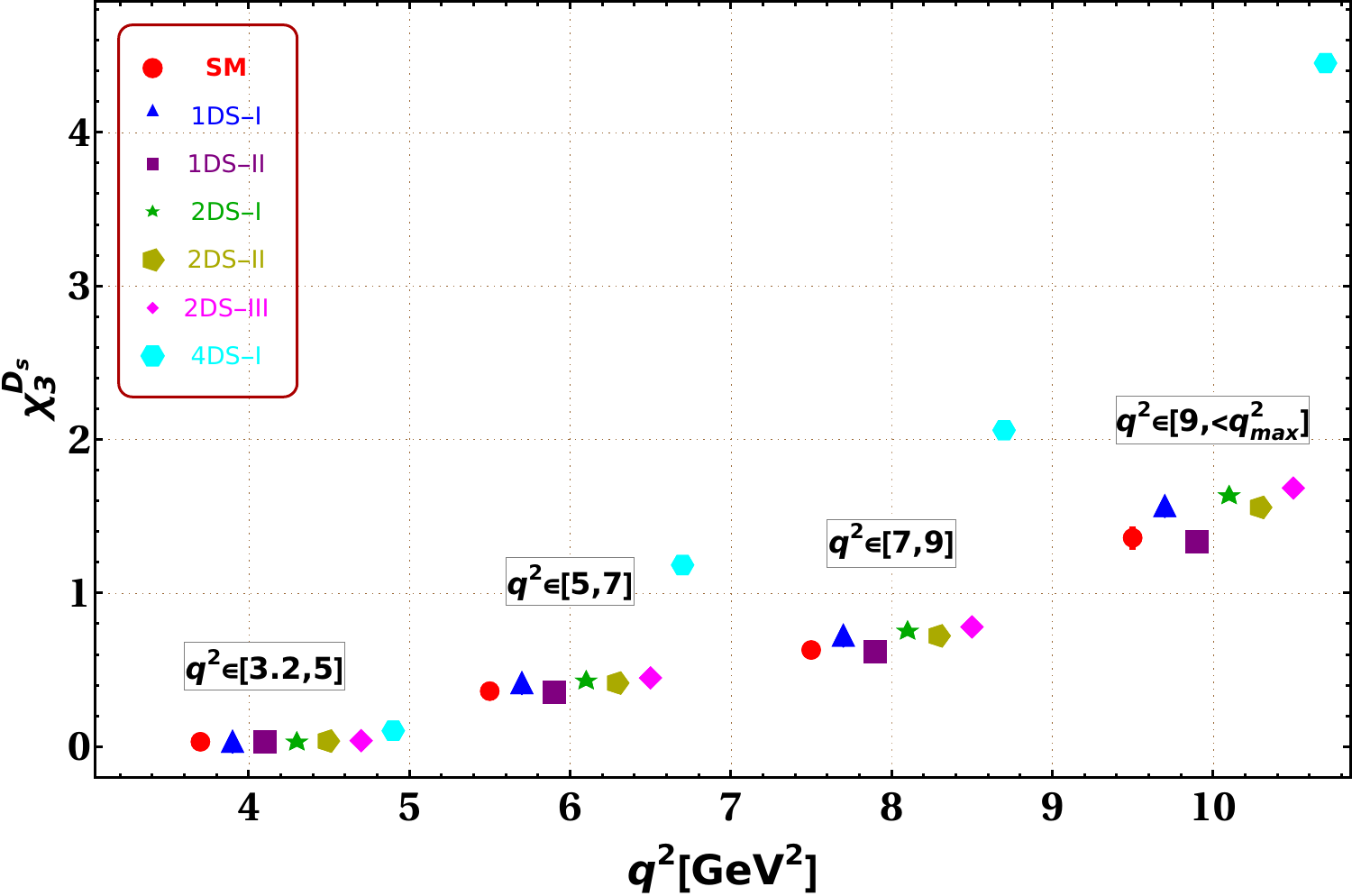} \quad
     \includegraphics[width=0.48\linewidth]{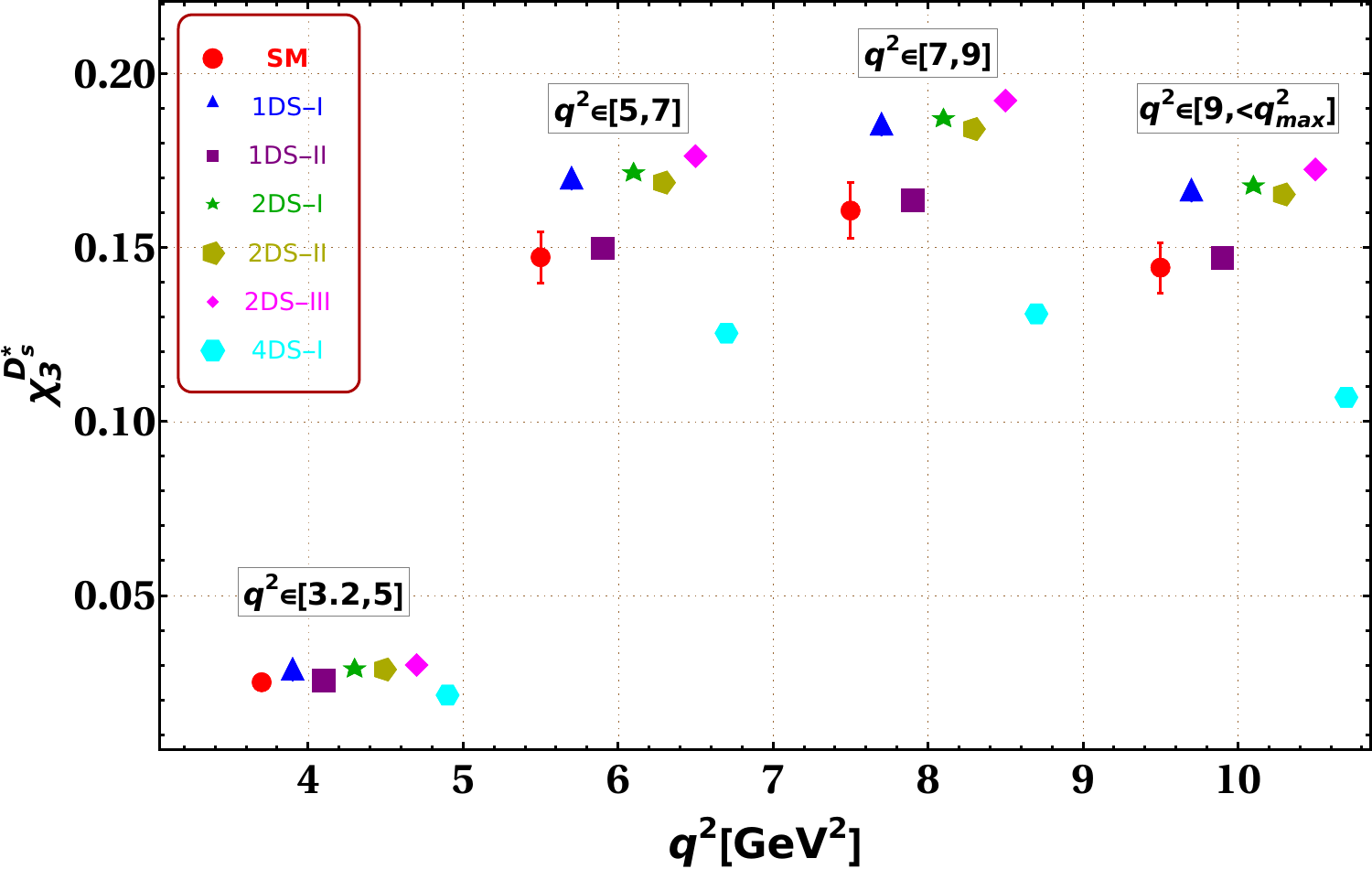}
     \includegraphics[width=0.48\linewidth]{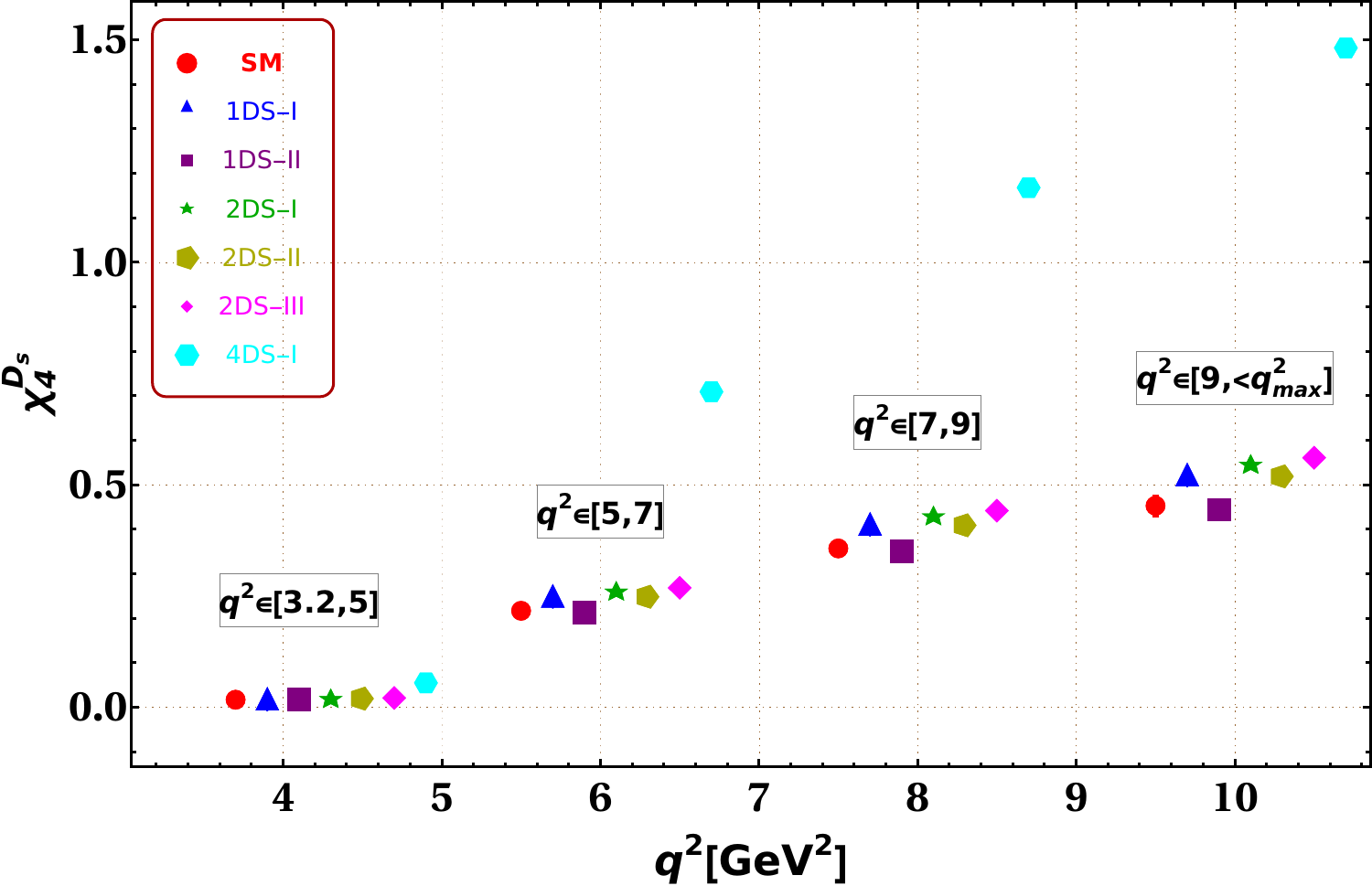} \quad
      \includegraphics[width=0.48\linewidth]{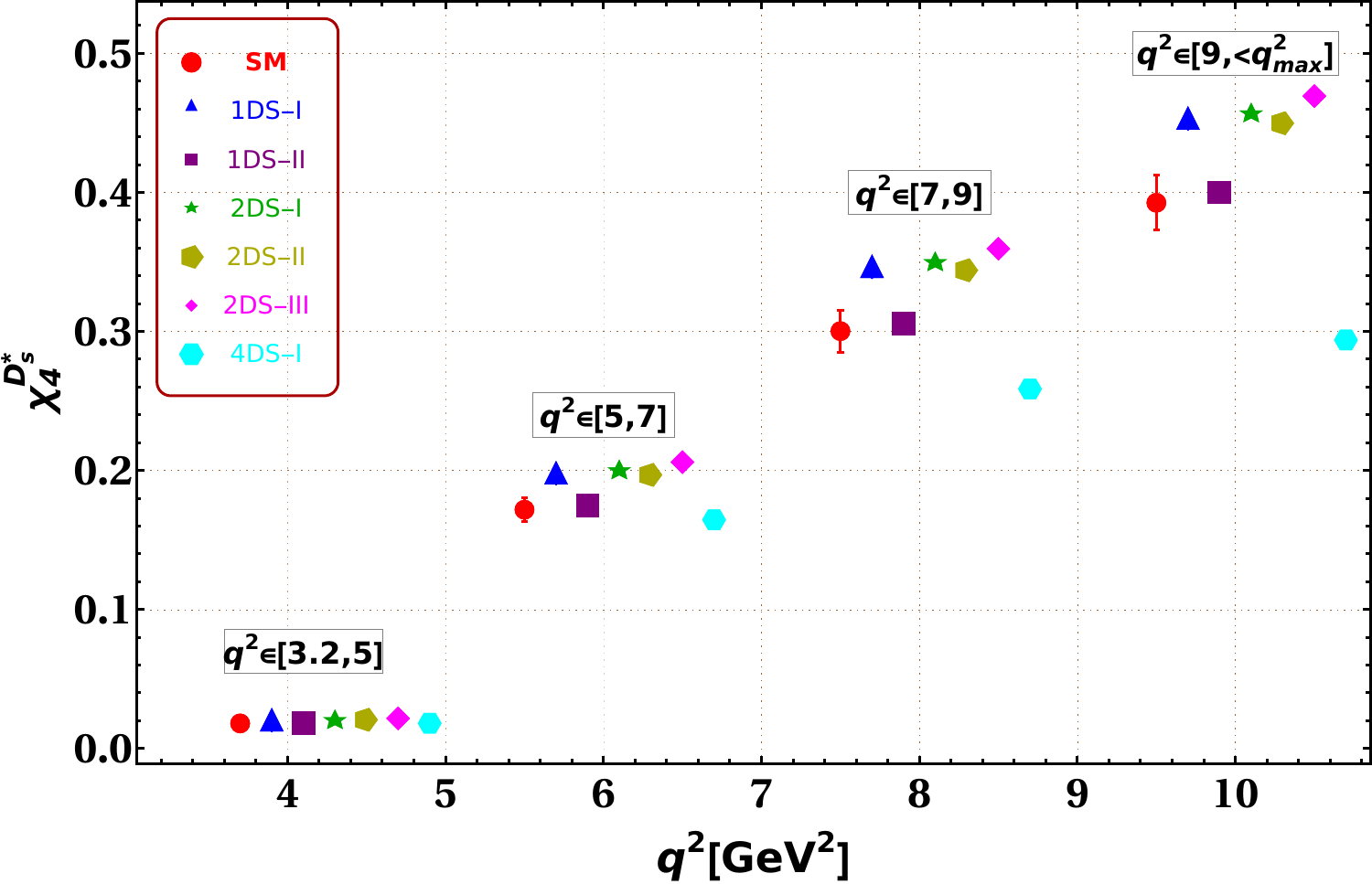}
    \caption{Same as Fig. \ref{fig:BR} for $P_\tau^{D_s}$ (top left panel), $P_\tau^{D_s^*}$ (top right panel), $\chi_3^{D_s}$ (middle left panel), $\chi_3^{D_s^*}$ (middle right panel), $\chi_4^{D_s}$ (bottom left panel) and  $\chi_4^{D_s^*}$ (bottom right panel).}
    \label{fig:Ptau}
\end{figure}

\begin{figure}[htb]
    \centering
    \includegraphics[width=0.48\linewidth]{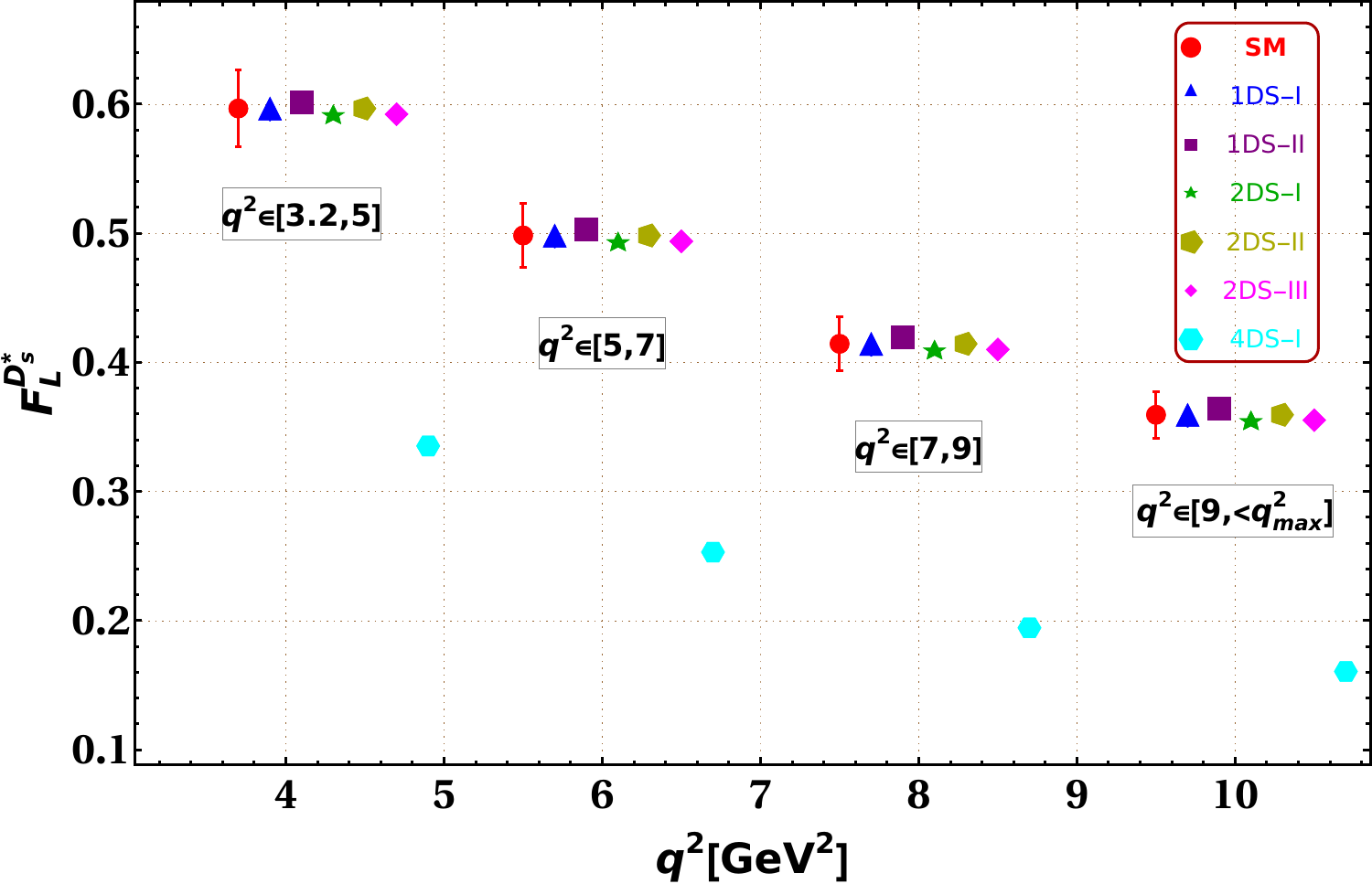} \quad
   \includegraphics[width=0.48\linewidth]{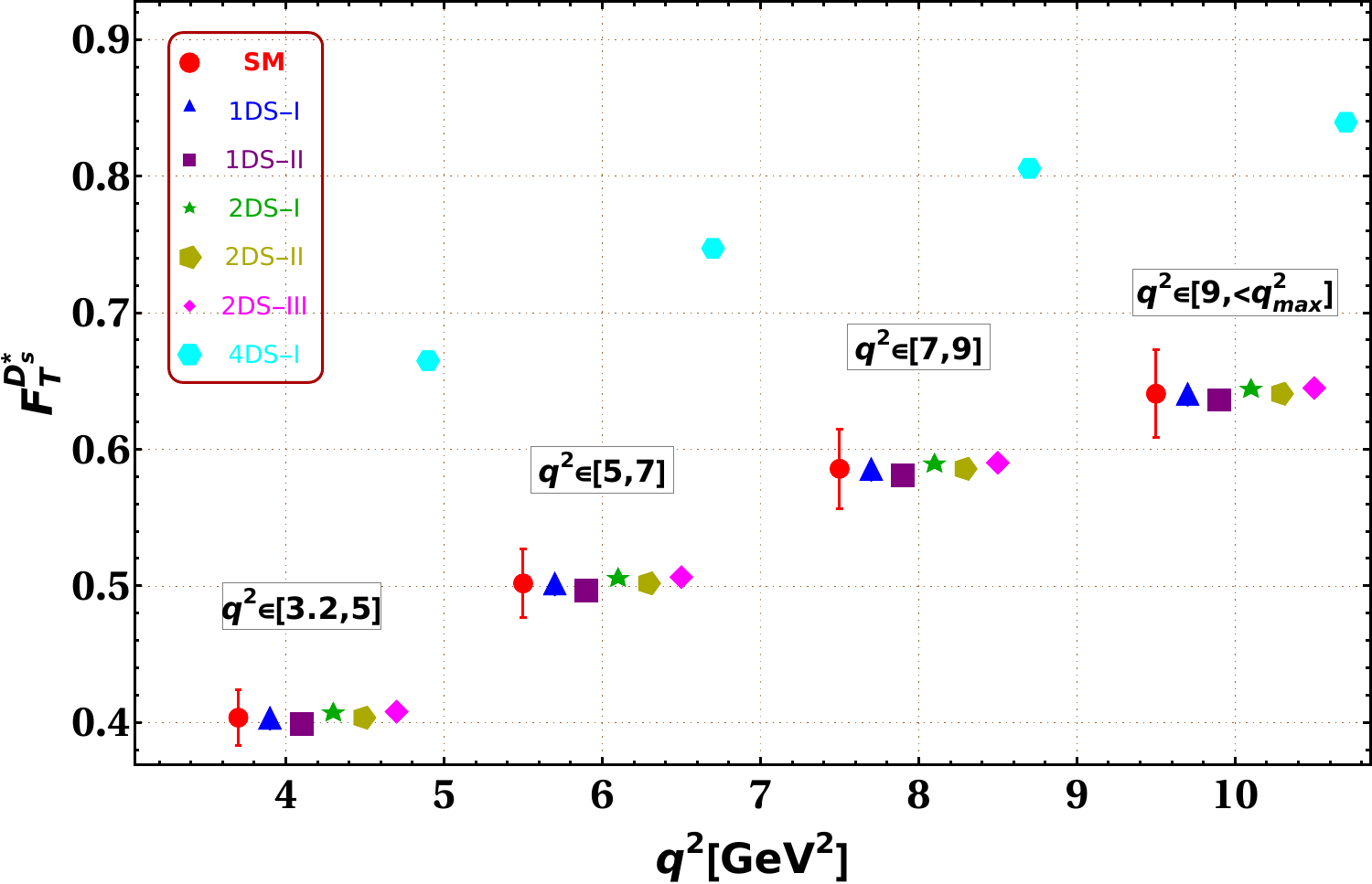} 
    \includegraphics[width=0.48\linewidth]{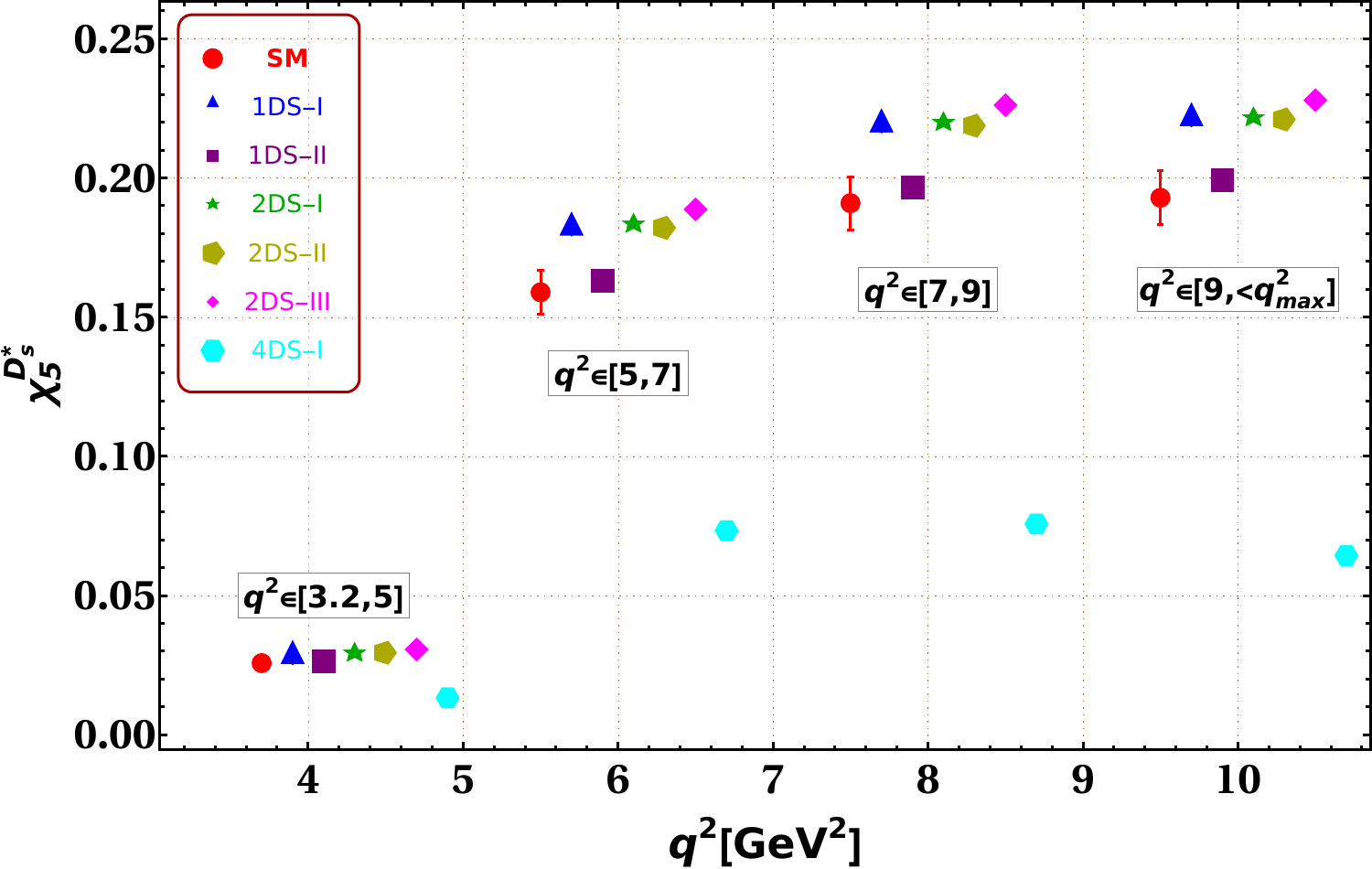} \quad 
     \includegraphics[width=0.48\linewidth]{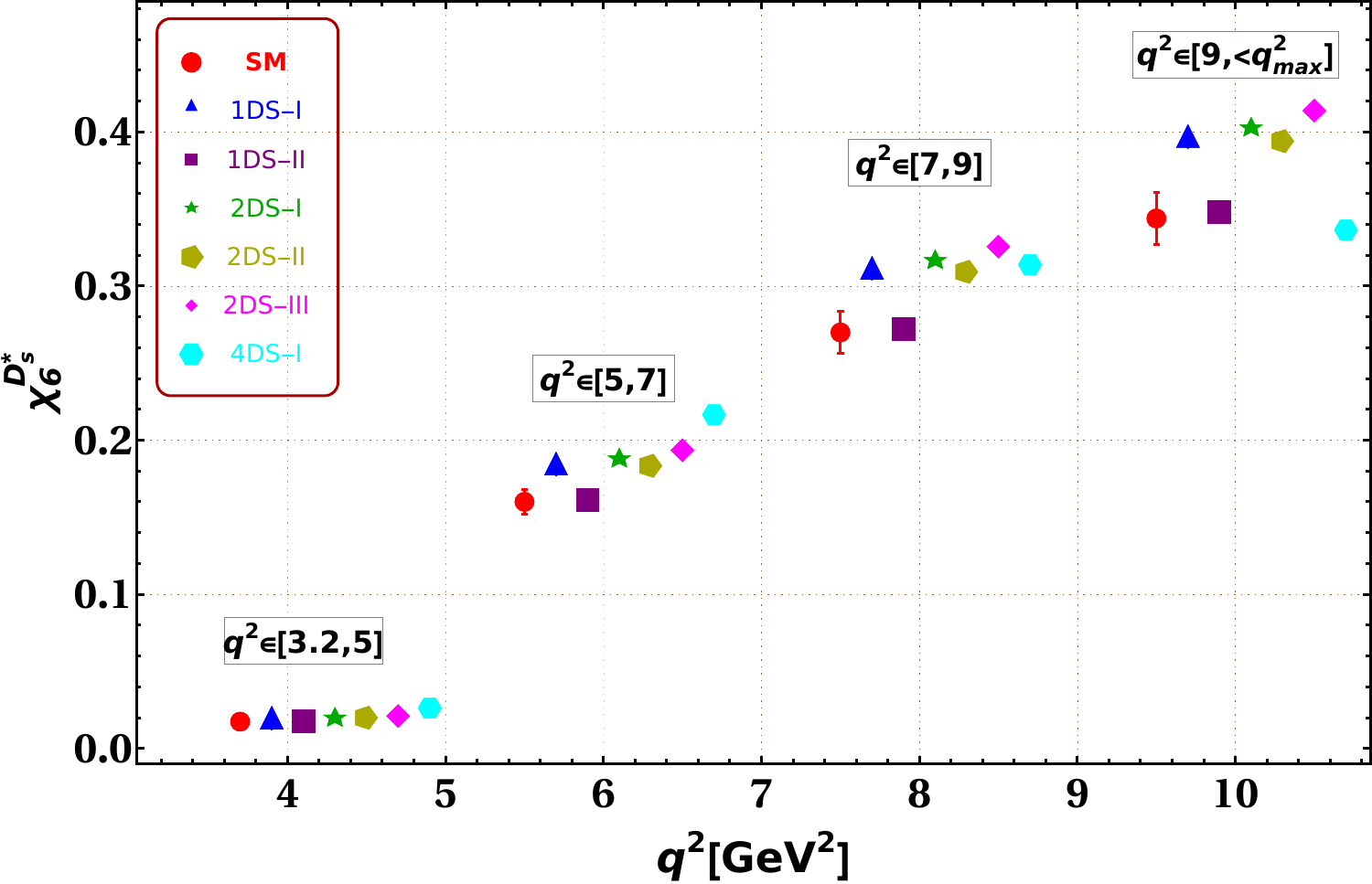}
    \caption{Same as Fig. \ref{fig:BR} for $F_L^{D_s^*}$ (top left panel), $F_T^{D_s^*}$ (top right panel), $\chi_5^{D_s^*}$ (bottom left panel) and  $\chi_6^{D_s^*}$ (bottom right panel).}
    \label{fig:long_Trans_C56}
\end{figure}

After collecting all the expressions for the branching ratios and key observables of the $B_s \to D_s^{(*)} \tau \bar{\nu}_\tau$ decay modes, we now proceed with the numerical analysis. For this analysis, we have taken the necessary input parameters from the PDG \cite{ParticleDataGroup:2024cfk} and the form factors for both $B_s \to D_s$ and $B_s \to D_s^*$ computed using the Lattice QCD method from \cite{McLean:2019qcx} and \cite{Harrison:2021tol}, respectively. Using the best-fit values from all the new physics scenarios (1D, 2D, and 4D), we computed the branching ratios, lepton non-universality, forward-backward asymmetry, and $\chi_{1,\ldots, 6}^{D_s^{(*)}}$ observables in four different $q^2$ bins: $q^2 \in [3.2, 5]~{\rm GeV}^2$, $q^2 \in [5, 7]~{\rm GeV}^2$, $q^2 \in [7, 9]~{\rm GeV}^2$, and $q^2 \in [9, q^2_{\rm max}]~{\rm GeV}^2$. The left (right) panel of Fig. \ref{fig:BR} depicts the branching ratio BR($B_s \to D_s^{(*)} \tau \bar{\nu}_\tau$) in all the new physics scenarios. The lepton non-universality parameters $R_{D_s}$ (left panel) and $R_{D_s^*}$ (right panel) are shown in Fig. \ref{fig:LNU}. The $q^2$ binwise predictions for the forward-backward asymmetries, $A_{FB}^{D_s}$ (top left panel) and $A_{FB}^{D_s^*}$ (top right panel), as well as the observables $\chi_{1,2}^{D_s}$ (left panel) and $\chi_{1,2}^{D_s^*}$ (right panel) are presented in the middle and bottom panels of Fig. \ref{fig:AFB}, respectively. Fig. \ref{fig:Ptau} represents similar results as Fig. \ref{fig:AFB} for the $\tau$ polarization asymmetries $P_\tau^{D_s^{(*)}}$ and the observables $\chi_{3,4}^{D_s^*}$. The longitudinal (transverse) polarization asymmetry of $D^*_s$ is presented in the top left (top right) panel of Fig. \ref{fig:long_Trans_C56}, while the bottom left (bottom right) panel represents the observables $\chi_{5,6}^{D_s^*}$.
%====================================================================
%=======================================================================
%====================================================================
%=======================================================================
Here, the red circle (error line) represents the central values of the SM along with the $1\sigma$ theoretical uncertainties. The predictions from the best-fit values of all the scenarios are presented in different graphics and colors: 
1DS-I $\rightarrow$ Blue, 2DS-I $\rightarrow$ Purple, 2DS-II $\rightarrow$ Green, 2DS-III $\rightarrow$ Dark Yellow,  2DS-IV $\rightarrow$ Magenta,  and 4DS-I $\rightarrow$ Cyan. The numerical values of the branching ratios and other physical observables for the decay $B_s \to D_s \tau \bar{\nu}_\tau$ in the SM and all the new physics scenarios across the four different $q^2$ bins are presented in Table~\ref{Tab:BstoDs_Obs} of Appendix~D. Tables~\ref{Tab:BstoDsstar_Obs} and~\ref{Tab:BstoDsstar_Chis} of Appendix~D include the numerical values of the observables for the process $B_s \to D_s^* \tau \bar{\nu}_\tau$.

%====================================================================
%=======================================================================
%====================================================================
%=======================================================================
%====================================================================
%========================================================================

By analyzing the physical observables in the SM and in the presence of various new physics scenarios, we observed the following:

\begin{itemize}
    \item \textbf{Branching Ratio:} Significant deviations from the SM predictions have been observed for the branching ratios of the decay processes $B_s \to D_s^{(*)} \tau \bar \nu_\tau$ in the $q^2$ bins $[5, 7]$ and $[7, 9]$, particularly for the 2DS-III and 4DS-I scenarios.
    
    \item \textbf{Lepton Non-universality:} The new physics contributions in the first $q^2$ bin show no effect. However, the 2DS-III and 4DS-I scenarios exhibit significant effects on the lepton non-universality parameters in the other $q^2$ bins.
    
    \item \textbf{Forward Backward Asymmetry:} There is no deviation in the forward backward asymmetry of the decay modes $B_s \to D_s \tau \bar \nu_\tau$. In contrast, significant deviations are found for the decay process $B_s \to D_s^* \tau \bar \nu_\tau$ across all $q^2$ bins. The contributions from the 2DS-I scenario are the most pronounced, followed by the 4DS-I scenario. The forward backward asymmetry is positive for both the 2DS-I and 4DS-I scenarios, and the zero-crossing of $A_{FB}^{D_s^*}$ is shifted to the third bin for the 2DS-III scenario.
    
    \item \textbf{$\boldsymbol{\tau}$ Polarization Asymmetry:} No deviations have been found in the $P_\tau^{D_s^{(*)}}$ observables due to the presence of new physics scenarios, except for a marginal deviation in the first three $q^2$ bins of $P_\tau^{D_s^*}$, as predicted by the 4DS-I scenario.

    \item \textbf{$\boldsymbol{D_s^*}$ Polarization Asymmetry:} The contributions from the new physics coefficients to the longitudinal and transverse polarization asymmetries are very marginal. 

\item \textbf{Observables $\boldsymbol{\chi_{1,\cdots, 6}^{D_s^{(*)}}}$:} 
The observables $\chi_{1,2}^{D_s}$ exhibit deviations due to the 4DS-I scenario. The observable $\chi_{1}^{D_s^*}$ shows significant effects from the 2DS-I and 4DS-I scenarios, while $\chi_{2}^{D_s^*}$ is notably affected by the 1DS-I, 2DS-II, and 4DS-I scenarios. The first bin of the observables $\chi_{3,4}^{D_s^{(*)}}$ shows no deviations with the inclusion of new physics. However, all the remaining bins of $\chi_{3,4}^{D_s^{(*)}}$ receive significant contributions from the 4DS-I scenario. Except for 1DS-II, all other new physics scenarios exhibit substantial deviations from the SM in the second, third, and fourth $q^2$ bins of the $\chi_3^{D_s^*}$ observable. Additionally, the observables $\chi_4^{D_s^{(*)}}$ show deviations in the third and fourth bins due to the influence of new physics. The 4DS-I scenario shows a large deviation from the SM for the observable $\chi_5^{D_s^*}$, whereas the deviation for $\chi_6^{D_s^*}$ is marginal.
These patterns indicate that the $\chi_i^{D_s^{(*)}}$ observables are particularly sensitive to shape distortions in the differential decay rates, which are often missed by normalized observables. This enhanced sensitivity makes them valuable probes for new physics.

In contrast to the $\chi_i^{D_s^{(*)}}$ observables,  the angular observable \(A_{FB}(q^2)\) and the polarization \(P_\tau(q^2)\) are defined as normalized ratios of helicity amplitudes. In the presence of vector-like NP operators, contributions typically scale both numerator and denominator in a similar way, resulting in a near cancellation of NP effects. Consequently, these observables are relatively insensitive to moderate NP induced shifts. On the other hand, the $\chi_i^{D_s^{(*)}}$  observables constructed from binned deviations in the differential rates, are sensitive to the absolute changes in decay distributions. Even small shape distortions across \(q^2\) bins lead to noticeable deviations when aggregated. This behavior is evident in Figs.~\ref{fig:AFB} and \ref{fig:Ptau}, where the normalized observables remain close to SM predictions, while the $\chi_i^{D_s^{(*)}}$ distributions show significant variation under NP. 
\end{itemize}

%=====================================
\section{Conclusion}
In conclusion, we investigated the $b \to c \tau \bar{\nu}_\tau$ decays, specifically focusing on $B_s \to D_s^{(*)} \tau \bar{\nu}_\tau$, using a model independent approach. We examined the sensitivity of the (axial)vector coefficients on branching ratios and various physical observables across four different $q^2$ bins: $q^2 \in [3.2, 5]$, $[5, 7]$, $[7, 9]$, and $[9, q_{\rm max}^2]$. The physical observables considered include forward backward asymmetry, $\tau$ polarization asymmetry, $D_s^*$ polarization asymmetry, lepton non-universality parameters, and the $\chi_{1,2,3,4,5,6}^{D_s^{(*)}}$ observables. We performed the analysis for six new physics scenarios, exploring all combinations of real and complex (axial)vector Wilson coefficients. Our findings indicate that the 2DS-III and 4DS-I scenarios significantly impact all observables across nearly all four $q^2$ bins, with some exceptions in the first bin for certain observables. The scenario 2DS-I also shows a good shift from the SM predictions for a few observables like foward backward asymmetry and  $\chi_{1,2}$ observables of $B_s \to D_s^{*}$ decay modes.  However, the $D_s^*$ and $\tau$ polarization asymmetries for the $B_s \to D_s^{(*)} \tau \bar{\nu}_\tau$ decay modes show no deviations from the SM predictions. The behavior predicted by the 1DS-II scenario aligns with that of the SM across all observables. Based on our analysis, we recommend investigating potential violations of lepton universality in the $B_s \to D_s^{(*)} \tau \bar{\nu}_\tau$ decay processes at B-factories and the LHCb experiment.

%When we looked at the impact of new physics parameters on decays $B_s\to D_s^{(*)}\tau\bar{\nu}_\tau$, we observed the 2DS-III and 4DS-I scenarios had the most pronounced contribution to the LNU parameters. The $\chi^{D_s^{(*)}}_{1,2,3,4,5,6}$ observables had deviation due to the presence of NP scenarios with significant effects from the 2DS-III and 4DS-I scenarios. Other NP scenarios showed significant deviation from the SM in the second, third, and fourth $q^2$ bins of $\chi_3^{D_s^*}$ observable.  Beside this, $D^*$ and $\tau$ polarization asymmetry have very edge gaps with SM, which implies the validity of SM for this weak decay channel. It suggests looking very preciously for these decays at the experimental side. 
%===================================

%=====================================
\section*{Acknowledgment}
%==================================================
AKY acknowledges sincere thanks to the Government of India's DST-INSPIRE Fellowship division for their financial assistance (ID No. IF210687).

%============================
\appendix
%======================================

\section{Helicity amplitudes for $B_s\to D^{(*)}_s\tau\bar{\nu}_{\tau}$}\label{A}
\begin{itemize}
    \item $\boldsymbol{B_s\to D_s\tau\bar{\nu}_{\tau}}$\\
    The helicity amplitudes  $H_{0,t} (q^2)$ of $B_s \to D_s$ decay mode in terms of the form factors $f_{+,0}(q^2)$ are given as \cite{Sakaki:2013bfa}
    \begin{align}
    H_0(q^2)&=\sqrt{\frac{\lambda_{D_s}}{q^2}}\,f_+(q^2), \nn \\
    H_t(q^2)&=\frac{M_B^2-M_{D_s}^2}{\sqrt{q^2}}\,f_0(q^2).
    \end{align}
    \item $\boldsymbol{B_s\to D_s^*\tau\bar{\nu}_{\tau}}$\\
     The helicity amplitudes  $H_{V,\lambda} (q^2)$ ($\lambda=\pm, 0, t$) of $B_s \to D_s^*$ are given as  \cite{Sakaki:2013bfa}
    \begin{align}
        H_{V,\pm}(q^2)&=(M_B+M_{D^*_s})\,A_1(q^2)\mp\frac{\sqrt{\lambda_{D_s^*}}}{M_B+M_{D^*_s}}\,V(q^2), \nn \\
        H_{V,0}(q^2)&=\frac{M_B+M_{D^*_s}}{2\,M_{D_s^*}\,\sqrt{q^2}}\Bigg[-\big(M_B^2-M_{D_s^*}^2-q^2\big)\,A_1(q^2)+\frac{\lambda_{D_s^*}}{(M_B+M_{D_s^*})^2}\,A_2(q^2)\Bigg],\nn \\
        H_{V,t}(q^2)&=-\frac{\sqrt{\lambda_{D_s^*}}}{\sqrt{q^2}}\,A_0(q^2).
         \end{align}
\end{itemize}\section{$B_s \to D_s$ form factors (Lattice QCD)}\label{B}
For the $B_s\to D_s$ decay, the form factors $f_{0,+}(q^2)$ with the Bourrely-Caprini-Lellouch (BCL) parameterization are defined as \cite{Bourrely:2008za}
\begin{align}
     f_0(q^2)&=\frac{1}{\Big(1-\frac{q^2}{M_{B^0_c}^2}\Big)}\,\sum_{n=0}^{N-1}\,a_n\,z^n(q^2),\nn \\
     f_+(q^2)&=\frac{1}{\Big(1-\frac{q^2}{M_{B_c^*}^2}\Big)}\,\sum_{n=0}^{N-1}\,a_n\,\Big(z^n(q^2)-\frac{n}{N}\,(-1)^{n-N}\,z^N(q^2)\Big)
\end{align}
where the $z$ polynomial is expressed as 
\begin{equation*}
    z(q^2)=\frac{\sqrt{t_+ - q^2}-\sqrt{t_+ }}{\sqrt{t_+ - q^2}+\sqrt{t_+ }},
\end{equation*} 
Here, $t_+=(M_{B_s}+M_{D_s})^2$, $N=3$ and $M_{B_c^0}=6.704$ and $M_{B^*_c}=6.332$ are the physical pole mass \cite{McLean:2019qcx}.  Table \ref{tab:a_value_Ds} includes the numerical values of the $z$-expenssion coefficients  in the full kinematic $q^2$ region, as calculated using the lattice QCD approach \cite{McLean:2019qcx}.

\begin{table}[htb]
    \centering
    \begin{tabular}{|c|c|c|} \hline
    ~Form factor ~&~ $z$-Coefficients ~  & ~Mean value~\\ \hline
      $f_0$   & $a_0$ &0.66574\\
              & $a_1$ &-0.25944\\
              & $a_2$ &-0.10636\\ \hline
      $f_+$   & $a_0$ &0.66574\\
              & $a_1$ &-3.23599\\
              & $a_2$ &-0.07478 \\ \hline
    \end{tabular}
     \caption{Numerical value for $z$-coefficients for $B_s\to D_s$ decay in lattice QCD.}
    \label{tab:a_value_Ds}
\end{table}
\section{$B_s \to D_s^*$ form factors (Lattice QCD)}\label{C}
In Lattice QCD, the $q^2$ dependent form factors: $V(q^2)$ and $A_{0,1,2}(q^2)$, using the z-expansion are defined as \cite{Harrison:2021tol,McLean:2019sds,Hill:2006ub}
\begin{equation}
    F(q^2)=\frac{1}{P(q^2)}\,\sum_{n=0}^{3}\,a_n\,z^n\,\mathcal{N}_n,
\end{equation}
where 
\begin{eqnarray*}
&&P(q^2)=\prod_{\it{M_{pole}}}\,z(q^2,M^2_{pole})\,, \\
&&z(q^2,t_0)=\frac{\sqrt{t_+ - q^2}-\sqrt{t_+ - t_0}}{\sqrt{t_+ - q^2}+\sqrt{t_+ - t_0}},        
\end{eqnarray*}
with
\begin{equation*}
    t_0=(M_{B_s} - M_{D_s^*})^2,~~~~~~~~~
    t_+=(M_B + M_{D^*})^2, ~~~~~~~\mathcal{N}_n=1. 
\end{equation*}
Here, $P(q^2)$ is a pole function that includes poles resulting from $b\Bar{c}$ where $A_0(q^2)$ is constructed with $0^-$, $V(q^2)$ with $1^-$ and $1^+$ states for $A_{1,2}(q^2)$. The predicted masses for the $B_c$ pseudoscalar, vector, and axial vector states below the $BD^*$ threshold, which are incorporated into our pole factor, are presented in Table \ref{tab:pole-mass}. The $z$-expansion coefficient, $a_n$, is listed in Table \ref{tab:form_factor_value}.
\begin{table}[htb]
    \centering
    \begin{tabular}{ccc}
    \hline\hline
    $0^-/\text{GeV}$  & $1^-/\text{GeV}$  & $1^+/\text{GeV}$  \\ \hline 
       6.275  & 6.335 &6.745 \\
       6.872 & 6.926 & 6.75 \\
       7.25 & 7.02 & 7.15 \\
       & 7.28 & 7.15 \\ \hline
    \end{tabular}
    \caption{The physical masses of $B_c$ pseudoscalar, axial vector and vector below the threshold $BD^*$.}
    \label{tab:pole-mass}
\end{table}

\begin{table}[htb]
    \centering
    \begin{tabular}{c|c c c c} \hline 
         & $a_0$ & $a_1$ & $a_2$ & $a_3$ \\ \hline
        $A_0$ & $0.1047\pm0.0057$ & $-0.43\pm0.13$& $-0.10\pm0.96$& $-0.03\pm1.00$\\
        $A_1$ &$0.0552\pm0.0021$&$-0.010\pm0.054$&$-0.03\pm0.77$&$0.06\pm0.99$\\
        $A_2$ &$0.059\pm0.011$&$-0.11\pm0.22$&$-0.25\pm0.79$&$-0.05\pm1.00$\\
        $V$&$0.100\pm0.011$&$-0.18\pm0.27$&$-0.006\pm0.998$&$0.0\pm1.0$ \\ \hline
    \end{tabular}
    \caption{The values of $z$-expansion coefficients $(a_n)$ for the vector, axial-vector and pseudoscalar form factors for $B_s\to D^*_s$ decay.}
    \label{tab:form_factor_value}
\end{table}

\section{Numerical Predictions}

%========================================================================
\begin{table}[htbp]
%\flushleft \flushleft \small
\centering
\setlength{\tabcolsep}{10pt}
\renewcommand{\arraystretch}{0.9} 
%\caption{ Numerical predictions for observables ($BR$,$R_{D^*_s}$,$A_{FB}^{D_s^*}$,$P_{\tau}^{D_s^*}$, $F_L^{D_s^*}$ and $F_T^{D_s^*}$) for $B_s\to D^*_s\tau\bar{\nu_{\tau}} $ in $q^2$ bins}\label{Tab:BctoDsstar_Obs}
\begin{tabular}{|l | l | l | l | l | l | l | l | l |}
\toprule[1.2pt] 
%\multicolumn{2}{|c|c|}{\textbf{Scenarios} & \textbf{Observables}} \\ \hline
\textbf{Scenarios}~ & \textbf{$\boldsymbol{BR\times 10^3}$} & ~\textbf{$\boldsymbol{R_{D_s}}$} ~&  \textbf{$\boldsymbol{A_{FB}^{D_s}}$} &\textbf{$\boldsymbol{P_{\tau}^{D_s}}$} &\textbf{$\boldsymbol{\chi_1^{D_s}}$} & ~\textbf{$\boldsymbol{\chi_2^{D_s}}$} ~&  \textbf{$\boldsymbol{\chi_3^{D_s}}$} &\textbf{$\boldsymbol{\chi_4^{D_s}}$} \\ \hline
\multicolumn{9}{|c|}{\textbf{$\boldsymbol{q^2\in[3.2-5]~{\rm GeV}^2}$}} \\ 
\hline
SM & 0.796&0.048&0.443& 0.310&0.035&0.013&0.032&0.017\\
1DS-I& 0.920 &0.056&0.443& 0.310&0.039&0.015&0.036&0.019\\
1DS-II&0.780 &0.047&0.443& 0.310&0.034&0.013&0.031&0.016\\
2DS-I& 0.962& 0.058&0.443& 0.310&0.042&0.016&0.038&0.020\\
2DS-II& 0.912&0.055&0.443& 0.310&0.039&0.015&0.036&0.019\\
2DS-III & 0.986&0.060&0.443& 0.310&0.043&0.017&0.039&0.021\\
4DS-I & 2.606&0.158&0.443& 0.310&0.114&0.044&0.103&0.054\\
\hline
\multicolumn{9}{|c|}{\textbf{$\boldsymbol{q^2\in[5-7]~{\rm GeV}^2}$}} \\ 
\hline
SM & 2.384&0.578&0.394& 0.251&0.403&0.175&0.361&0.216\\
1DS-I& 2.755& 0.668&0.394& 0.251&0.465&0.202&0.417&0.250\\
1DS-II& 2.336&0.566&0.394& 0.251&0.395&0.172&0.354&0.212\\
2DS-I&2.879& 0.698&0.394& 0.251&0.486&0.211&0.436&0.261\\
2DS-II& 2.731&0.662&0.394& 0.251&0.461&0.201&0.414&0.248\\
2DS-III &2.952&0.715&0.394& 0.251&0.499&0.217&0.447&0.268\\
4DS-I &7.804  &1.891&0.394& 0.251&1.318&0.573&1.183&0.709\\
\hline
\multicolumn{9}{|c|}{\textbf{$\boldsymbol{q^2\in[7-9]~{\rm GeV}^2}$}} \\ 
\hline
SM & 2.410&0.986&0.349& 0.277&0.665&0.321&0.629&0.357\\
1DS-I& 2.790& 1.139&0.349& 0.277&0.769&0.371&0.727&0.412\\
1DS-II&2.366& 0.966&0.349& 0.277&0.652&0.314&0.617&0.349\\
2DS-I& 2.916&1.191&0.349& 0.277&0.803&0.387&0.760& 0.431\\
2DS-II& 2.767&1.130&0.349& 0.277&0.762&0.367&0.721&0.409\\
2DS-III & 2.990&1.221&0.349& 0.277&0.824&0.397&0.779&0.441\\
4DS-I & 7.905&3.228&0.349& 0.277&2.178&1.050&2.060&1.167\\
\hline
\multicolumn{9}{|c|}{\textbf{$\boldsymbol{q^2\in[9-q^2_{max}]~{\rm GeV}^2}$}} \\ 
\hline
SM & 1.576&1.812& 0.285&0.501& 1.164&0.649&1.359&0.452\\
1DS-I&1.821&2.094&0.285& 0.501&1.345&0.749&1.571&0.523\\
1DS-II&1.545& 1.776&0.285& 0.501&1.141&0.635&1.332&0.443\\
2DS-I&1.904& 2.188&0.285& 0.501&1.406&0.783&1.642&0.547\\
2DS-II&1.806&2.076&0.285& 0.501&1.334&0.743&1.558&0.518\\
2DS-III &1.952&2.244&0.285& 0.501&1.441&0.803&1.683&0.560\\
4DS-I &5.160&5.932&0.285& 0.501&3.810&2.122&4.451&1.481\\
\bottomrule[1.2pt] 
\end{tabular}
\caption{Numerical predictions for $B_s \to D_s \tau \bar{\nu}_{\tau}$ observables in various $q^2$ bins.}\label{Tab:BstoDs_Obs}
\end{table}
%=======================================================================e
\begin{table}[htbp]
%\flushleft \flushleft \small
\centering
\setlength{\tabcolsep}{11pt}
\renewcommand{\arraystretch}{0.9} 
%\caption{ Numerical predictions for observables ($BR$,$R_{D^*_s}$,$A_{FB}^{D_s^*}$,$P_{\tau}^{D_s^*}$, $F_L^{D_s^*}$ and $F_T^{D_s^*}$) for $B_s\to D^*_s\tau\bar{\nu_{\tau}} $ in $q^2$ bins}\label{Tab:BctoDsstar_Obs}
\begin{tabular}{|l | l | l | l | l | l | l |}
\toprule[1.2pt] 
%\multicolumn{2}{|c|c|}{\textbf{Scenarios} & \textbf{Observables}} \\ \hline
\textbf{Scenarios}~ & \textbf{$\boldsymbol{BR\times 10^3}$} & ~\textbf{$\boldsymbol{R_{D^*_s}}$} ~&  \textbf{$\boldsymbol{A_{FB}^{D_s^*}}$} &\textbf{$\boldsymbol{P_{\tau}^{D_s^*}}$} & $\boldsymbol{F_L^{D_s^*}}$ & $\boldsymbol{F_T^{D_s^*}}$ \\ \hline
\multicolumn{7}{|c|}{\textbf{$\boldsymbol{q^2\in[3.2-5]~{\rm GeV}^2}$}} \\ 
\hline
SM & $1.060$ & 0.043& 0.062& 0.164&0.597&0.403\\
1DS-I&$1.225$ & 0.049&0.062& 0.164&0.597&0.403\\
1DS-II&$1.078$ & 0.044& 0.066& 0.165&0.601&0.399\\
2DS-I& $1.239$ & 0.050&0.441 &0.163&0.592&0.408 \\
2DS-II& $1.215$ & 0.049& 0.062& 0.164&0.597&0.403\\
2DS-III & $1.271$& 0.052&0.126&0.163&0.592&0.408 \\
4DS-I & $0.971$ & 0.040&0.216& 0.083&0.335&0.665\\
\hline
\multicolumn{7}{|c|}{\textbf{$\boldsymbol{q^2\in[5-7]~{\rm GeV}^2}$}} \\ \hline
SM & $4.053$ & 0.319&-0.063&-0.077&0.498&0.502 \\
1DS-I&$4.684$ & 0.369& -0.063& -0.077&0.498&0.502 \\
1DS-II&$4.120$ &0.324& -0.058& -0.076&0.503&0.497\\
2DS-I& $4.735$ & 0.373& 0.399& -0.078&0.494&0.506\\
2DS-II& $4.644$ & 0.365& -0.063&-0.077&0.498&0.502 \\
2DS-III & $4.857$ & 0.382&0.016&-0.078&0.494&0.506 \\
4DS-I & $3.684$ & 0.290&0.163& -0.135&0.253&0.747 \\
\hline
\multicolumn{7}{|c|}{\textbf{$\boldsymbol{q^2\in[7-9]~{\rm GeV}^2}$}} \\ \hline
SM & $5.465$ & 0.461&-0.132& -0.303&0.414&0.586\\
1DS-I&$6.314$ & 0.533& -0.132&-0.303&0.414&0.586 \\
1DS-II&$5.559$ & 0.469& -0.128& -0.303&0.419&0.581\\
2DS-I& $6.378$ & 0.538& 0.333& -0.303&0.410&0.590\\
2DS-II& $6.261$&0.528& -0.132& -0.303&0.414& 0.586\\
2DS-III & $6.543$&0.552& -0.053&-0.303&0.410& 0.590\\
4DS-I & $4.619$&0.390& 0.122&-0.327&0.194&0.806 \\
\hline
\multicolumn{7}{|c|}{\textbf{$\boldsymbol{q^2\in[9-q^2_{max}]~{\rm GeV}^2}$ }} \\ \hline
SM & $3.213$ & 0.537&-0.115& -0.463&0.359&0.641\\
1DS-I&$3.713$ &0.620 &-0.115&-0.463&0.359&0.641 \\
1DS-II&$3.274$ &0.547 &-0.112&-0.463&0.364&0.636 \\
2DS-I& $3.745$ &0.626&0.215& -0.463&0.355&0.645\\
2DS-II& $3.681$&0.615& -0.115&-0.463&0.359&0.641\\
2DS-III & $3.841$&0.642&-0.059& -0.463&0.355&0.645\\
4DS-I & $2.399$&0.401&  0.078& -0.466&0.161&0.839\\ 
\bottomrule[1.2pt] 
\end{tabular}
\caption{ Numerical predictions for $B_s\to D^*_s\tau\bar{\nu}_{\tau}$ observables in $q^2$ bins.}\label{Tab:BstoDsstar_Obs}
\end{table}

%--------------------==========

\begin{table}[htbp]
%\flushleft \flushleft 
%\small
\centering
\setlength{\tabcolsep}{11pt}
\renewcommand{\arraystretch}{0.9} 
%\caption{ Numerical predictions for observables ($BR$,$R_{D^*_s}$,$A_{FB}^{D_s^*}$,$P_{\tau}^{D_s^*}$, $F_L^{D_s^*}$ and $F_T^{D_s^*}$) for $B_s\to D^*_s\tau\bar{\nu_{\tau}} $ in $q^2$ bins}\label{Tab:BctoDsstar_Obs}
\begin{tabular}{|l | l | l | l | l | l | l |}
\toprule[1.2pt] 
%\multicolumn{2}{|c|c|}{\textbf{Scenarios} & \textbf{Observables}} \\ \hline
\textbf{Scenarios}~ & \textbf{$\boldsymbol{\chi_1^{D_s^*}}$} & ~\textbf{$\boldsymbol{\chi_2^{D_s^*}}$} ~&  \textbf{$\boldsymbol{\chi_3^{D_s^*}}$} &\textbf{$\boldsymbol{\chi_4^{D_s^*}}$} & \textbf{$\boldsymbol{\chi_5^{D_s^*}}$}& \textbf{$\boldsymbol{\chi_6^{D_s^*}}$} \\ \hline
\multicolumn{7}{|c|}{\textbf{$\boldsymbol{q^2\in[3.2-5]~{\rm GeV}^2}$}} \\ 
\hline
SM & 0.023 &0.020 &0.020&0.018&0.026& 0.017 \\
1DS-I& 0.026&0.023 &0.029&0.021& 0.030&0.020 \\
1DS-II&0.023 &0.020 &0.025&0.018&0.026& 0.017 \\
2DS-I&0.036  & 0.014 & 0.029& 0.021&0.029&0.021 \\
2DS-II& 0.026 &0.023 &0.029& 0.021&0.029& 0.020\\
2DS-III &0.029 &0.023 &0.030 &0.022& 0.031& 0.021\\
4DS-I & 0.024 &0.015 & 0.021& 0.018&0.013& 0.026\\
\hline
\multicolumn{7}{|c|}{\textbf{$\boldsymbol{q^2\in[5-7]~{\rm GeV}^2}$}} \\ 
\hline
SM & 0.149 & 0.169& 0.147&0.172&0.159 &0.160\\
1DS-I&0.173 & 0.196& 0.170&0.198&0.184& 0.185\\
1DS-II&0.153 & 0.172& 0.149& 0.174&0.163&0.161 \\
2DS-I&0.261  &0.112  &0.172&0.201&0.184 & 0.189\\
2DS-II& 0.171  &0.194 &0.169&0.197& 0.182& 0.183\\
2DS-III & 0.194 & 0.188& 0.176&0.206& 0.189& 0.193\\
4DS-I & 0.169  & 0.121& 0.125& 0.164&0.073 & 0.217\\
\hline
\multicolumn{7}{|c|}{\textbf{$\boldsymbol{q^2\in[7-9]~{\rm GeV}^2}$}} \\ 
\hline
SM & 0.199 & 0.261 &0.161& 0.300&0.191& 0.269\\
1DS-I& 0.231 & 0.301&0.186&0.347&0.221 & 0.312\\
1DS-II& 0.204& 0.264 &0.164&0.305&0.197 & 0.272\\
2DS-I& 0.358  & 0.179 &0.187&0.350& 0.220& 0.317\\
2DS-II&0.229  & 0.299&0.184& 0.344&0.219& 0.309\\
2DS-III &0.261 & 0.290 &0.192& 0.360&0.226& 0.326\\
4DS-I &0.218  & 0.171&0.131&0.259&0.076& 0.314 \\
\hline
\multicolumn{7}{|c|}{\textbf{$\boldsymbol{q^2\in[9-q^2_{max}]~{\rm GeV}^2}$}} \\ 
\hline
SM &0.237 & 0.299&0.144&0.393& 0.193 & 0.344\\
1DS-I&0.274 & 0.346&0.167&0.454&0.223 & 0.397\\
1DS-II& 0.243 & 0.304&0.147& 0.399& 0.199& 0.348\\
2DS-I& 0.380 & 0.246 &0.168&0.457&0.222&  0.403\\
2DS-II& 0.272 & 0.343&0.165&0.449&0.221 & 0.394\\
2DS-III & 0.302 & 0.340 &0.172&0.469&0.228&  0.414\\
4DS-I & 0.216   & 0.185&0.107 &0.294&0.064&  0.336\\
\bottomrule[1.2pt] 
\end{tabular}
\caption{ Numerical predictions for observables of $B_s\to D^*_s\tau\bar{\nu}_{\tau} $ in $q^2$ bins}\label{Tab:BstoDsstar_Chis}
\end{table}
%=====================

%\bibliographystyle{JHEP}
\bibliography{referance}

\begin{thebibliography}{62}
\expandafter\ifx\csname natexlab\endcsname\relax\def\natexlab#1{#1}\fi
\expandafter\ifx\csname bibnamefont\endcsname\relax
  \def\bibnamefont#1{#1}\fi
\expandafter\ifx\csname bibfnamefont\endcsname\relax
  \def\bibfnamefont#1{#1}\fi
\expandafter\ifx\csname citenamefont\endcsname\relax
  \def\citenamefont#1{#1}\fi
\expandafter\ifx\csname url\endcsname\relax
  \def\url#1{\texttt{#1}}\fi
\expandafter\ifx\csname urlprefix\endcsname\relax\def\urlprefix{URL }\fi
\providecommand{\bibinfo}[2]{#2}
\providecommand{\eprint}[2][]{\url{#2}}

\bibitem[{\citenamefont{et~al}(1997)}]{BUSKULIC1997373}
\bibinfo{author}{\bibfnamefont{D.~B.} \bibnamefont{et~al}},
  \bibinfo{journal}{Physics Letters B} \textbf{\bibinfo{volume}{395}},
  \bibinfo{pages}{373} (\bibinfo{year}{1997}), ISSN \bibinfo{issn}{0370-2693},
  \urlprefix\url{https://www.sciencedirect.com/science/article/pii/S0370269397000713}.

\bibitem[{\citenamefont{Bartelt et~al.}(1999)}]{CLEO:1998qvx}
\bibinfo{author}{\bibfnamefont{J.~E.} \bibnamefont{Bartelt}}
  \bibnamefont{et~al.} (\bibinfo{collaboration}{CLEO}), \bibinfo{journal}{Phys.
  Rev. Lett.} \textbf{\bibinfo{volume}{82}}, \bibinfo{pages}{3746}
  (\bibinfo{year}{1999}), \eprint{hep-ex/9811042}.

\bibitem[{\citenamefont{Amhis et~al.}(2021)}]{HFLAV:2019otj}
\bibinfo{author}{\bibfnamefont{Y.~S.} \bibnamefont{Amhis}} \bibnamefont{et~al.}
  (\bibinfo{collaboration}{HFLAV}), \bibinfo{journal}{Eur. Phys. J. C}
  \textbf{\bibinfo{volume}{81}}, \bibinfo{pages}{226} (\bibinfo{year}{2021}),
  \eprint{1909.12524}.

\bibitem[{\citenamefont{Bernlochner et~al.}(2022)\citenamefont{Bernlochner,
  Sevilla, Robinson, and Wormser}}]{Bernlochner:2021vlv}
\bibinfo{author}{\bibfnamefont{F.~U.} \bibnamefont{Bernlochner}},
  \bibinfo{author}{\bibfnamefont{M.~F.} \bibnamefont{Sevilla}},
  \bibinfo{author}{\bibfnamefont{D.~J.} \bibnamefont{Robinson}},
  \bibnamefont{and} \bibinfo{author}{\bibfnamefont{G.}~\bibnamefont{Wormser}},
  \bibinfo{journal}{Rev. Mod. Phys.} \textbf{\bibinfo{volume}{94}},
  \bibinfo{pages}{015003} (\bibinfo{year}{2022}), \eprint{2101.08326}.

\bibitem[{\citenamefont{Blanke et~al.}(2019)\citenamefont{Blanke, Crivellin,
  de~Boer, Kitahara, Moscati, Nierste, and
  Ni\v{s}and\v{z}i\'c}}]{Blanke:2018yud}
\bibinfo{author}{\bibfnamefont{M.}~\bibnamefont{Blanke}},
  \bibinfo{author}{\bibfnamefont{A.}~\bibnamefont{Crivellin}},
  \bibinfo{author}{\bibfnamefont{S.}~\bibnamefont{de~Boer}},
  \bibinfo{author}{\bibfnamefont{T.}~\bibnamefont{Kitahara}},
  \bibinfo{author}{\bibfnamefont{M.}~\bibnamefont{Moscati}},
  \bibinfo{author}{\bibfnamefont{U.}~\bibnamefont{Nierste}}, \bibnamefont{and}
  \bibinfo{author}{\bibfnamefont{I.}~\bibnamefont{Ni\v{s}and\v{z}i\'c}},
  \bibinfo{journal}{Phys. Rev. D} \textbf{\bibinfo{volume}{99}},
  \bibinfo{pages}{075006} (\bibinfo{year}{2019}), \eprint{1811.09603}.

\bibitem[{\citenamefont{Fedele et~al.}(2023)\citenamefont{Fedele, Blanke,
  Crivellin, Iguro, Kitahara, Nierste, and Watanabe}}]{Fedele:2022iib}
\bibinfo{author}{\bibfnamefont{M.}~\bibnamefont{Fedele}},
  \bibinfo{author}{\bibfnamefont{M.}~\bibnamefont{Blanke}},
  \bibinfo{author}{\bibfnamefont{A.}~\bibnamefont{Crivellin}},
  \bibinfo{author}{\bibfnamefont{S.}~\bibnamefont{Iguro}},
  \bibinfo{author}{\bibfnamefont{T.}~\bibnamefont{Kitahara}},
  \bibinfo{author}{\bibfnamefont{U.}~\bibnamefont{Nierste}}, \bibnamefont{and}
  \bibinfo{author}{\bibfnamefont{R.}~\bibnamefont{Watanabe}},
  \bibinfo{journal}{Phys. Rev. D} \textbf{\bibinfo{volume}{107}},
  \bibinfo{pages}{055005} (\bibinfo{year}{2023}), \eprint{2211.14172}.

\bibitem[{\citenamefont{Dutta et~al.}(2013)\citenamefont{Dutta, Bhol, and
  Giri}}]{Dutta:2013qaa}
\bibinfo{author}{\bibfnamefont{R.}~\bibnamefont{Dutta}},
  \bibinfo{author}{\bibfnamefont{A.}~\bibnamefont{Bhol}}, \bibnamefont{and}
  \bibinfo{author}{\bibfnamefont{A.~K.} \bibnamefont{Giri}},
  \bibinfo{journal}{Phys. Rev. D} \textbf{\bibinfo{volume}{88}},
  \bibinfo{pages}{114023} (\bibinfo{year}{2013}), \eprint{1307.6653}.

\bibitem[{\citenamefont{Lees et~al.}(2012)}]{BaBar:2012obs}
\bibinfo{author}{\bibfnamefont{J.~P.} \bibnamefont{Lees}} \bibnamefont{et~al.}
  (\bibinfo{collaboration}{BaBar}), \bibinfo{journal}{Phys. Rev. Lett.}
  \textbf{\bibinfo{volume}{109}}, \bibinfo{pages}{101802}
  (\bibinfo{year}{2012}), \eprint{1205.5442}.

\bibitem[{\citenamefont{Lees et~al.}(2024)}]{BaBar:2023kug}
\bibinfo{author}{\bibfnamefont{J.~P.} \bibnamefont{Lees}} \bibnamefont{et~al.}
  (\bibinfo{collaboration}{BaBar}), \bibinfo{journal}{Phys. Rev. D}
  \textbf{\bibinfo{volume}{110}}, \bibinfo{pages}{032018}
  (\bibinfo{year}{2024}), \eprint{2311.15071}.

\bibitem[{\citenamefont{Huschle et~al.}(2015)}]{Belle:2015qfa}
\bibinfo{author}{\bibfnamefont{M.}~\bibnamefont{Huschle}} \bibnamefont{et~al.}
  (\bibinfo{collaboration}{Belle}), \bibinfo{journal}{Phys. Rev. D}
  \textbf{\bibinfo{volume}{92}}, \bibinfo{pages}{072014}
  (\bibinfo{year}{2015}), \eprint{1507.03233}.

\bibitem[{\citenamefont{Hirose et~al.}(2018)}]{Belle:2017ilt}
\bibinfo{author}{\bibfnamefont{S.}~\bibnamefont{Hirose}} \bibnamefont{et~al.}
  (\bibinfo{collaboration}{Belle}), \bibinfo{journal}{Phys. Rev. D}
  \textbf{\bibinfo{volume}{97}}, \bibinfo{pages}{012004}
  (\bibinfo{year}{2018}), \eprint{1709.00129}.

\bibitem[{\citenamefont{Caria et~al.}(2020)}]{Belle:2019rba}
\bibinfo{author}{\bibfnamefont{G.}~\bibnamefont{Caria}} \bibnamefont{et~al.}
  (\bibinfo{collaboration}{Belle}), \bibinfo{journal}{Phys. Rev. Lett.}
  \textbf{\bibinfo{volume}{124}}, \bibinfo{pages}{161803}
  (\bibinfo{year}{2020}), \eprint{1910.05864}.

\bibitem[{\citenamefont{Adachi et~al.}(2025)}]{Belle-II:2025yjp}
\bibinfo{author}{\bibfnamefont{I.}~\bibnamefont{Adachi}} \bibnamefont{et~al.}
  (\bibinfo{collaboration}{Belle-II}) (\bibinfo{year}{2025}),
  \eprint{2504.11220}.

\bibitem[{\citenamefont{Aaij et~al.}(2015)}]{LHCb:2015gmp}
\bibinfo{author}{\bibfnamefont{R.}~\bibnamefont{Aaij}} \bibnamefont{et~al.}
  (\bibinfo{collaboration}{LHCb}), \bibinfo{journal}{Phys. Rev. Lett.}
  \textbf{\bibinfo{volume}{115}}, \bibinfo{pages}{111803}
  (\bibinfo{year}{2015}), \bibinfo{note}{[Erratum: Phys.Rev.Lett. 115, 159901
  (2015)]}, \eprint{1506.08614}.

\bibitem[{\citenamefont{Aaij et~al.}(2018{\natexlab{a}})}]{LHCb:2017rln}
\bibinfo{author}{\bibfnamefont{R.}~\bibnamefont{Aaij}} \bibnamefont{et~al.}
  (\bibinfo{collaboration}{LHCb}), \bibinfo{journal}{Phys. Rev. D}
  \textbf{\bibinfo{volume}{97}}, \bibinfo{pages}{072013}
  (\bibinfo{year}{2018}{\natexlab{a}}), \eprint{1711.02505}.

\bibitem[{\citenamefont{Banerjee
  et~al.}(2024)}]{HeavyFlavorAveragingGroupHFLAV:2024ctg}
\bibinfo{author}{\bibfnamefont{S.}~\bibnamefont{Banerjee}} \bibnamefont{et~al.}
  (\bibinfo{collaboration}{Heavy Flavor Averaging Group (HFLAV)})
  (\bibinfo{year}{2024}), \eprint{2411.18639}.

\bibitem[{\citenamefont{Collaboration}(2025)}]{HFLAV2025}
\bibinfo{author}{\bibfnamefont{H.}~\bibnamefont{Collaboration}}
  (\bibinfo{collaboration}{HFLAV}), \emph{\bibinfo{title}{`` preliminary
  average of r(d) and r(d*) for spring 2025"}} (\bibinfo{year}{2025}),
  \urlprefix\url{https://hflav-eos.web.cern.ch/hflav-eos/semi/spring25/html/RDsDsstar/RDRDs.html}.

\bibitem[{\citenamefont{Aaij et~al.}(2018{\natexlab{b}})}]{LHCb:2017vlu}
\bibinfo{author}{\bibfnamefont{R.}~\bibnamefont{Aaij}} \bibnamefont{et~al.}
  (\bibinfo{collaboration}{LHCb}), \bibinfo{journal}{Phys. Rev. Lett.}
  \textbf{\bibinfo{volume}{120}}, \bibinfo{pages}{121801}
  (\bibinfo{year}{2018}{\natexlab{b}}), \eprint{1711.05623}.

\bibitem[{\citenamefont{Harrison et~al.}(2020)\citenamefont{Harrison, Davies,
  and Lytle}}]{Harrison:2020nrv}
\bibinfo{author}{\bibfnamefont{J.}~\bibnamefont{Harrison}},
  \bibinfo{author}{\bibfnamefont{C.~T.~H.} \bibnamefont{Davies}},
  \bibnamefont{and} \bibinfo{author}{\bibfnamefont{A.}~\bibnamefont{Lytle}}
  (\bibinfo{collaboration}{LATTICE-HPQCD}), \bibinfo{journal}{Phys. Rev. Lett.}
  \textbf{\bibinfo{volume}{125}}, \bibinfo{pages}{222003}
  (\bibinfo{year}{2020}), \eprint{2007.06956}.

\bibitem[{CMS(EPS-HEP 2023 Conference)}]{CMS_Collaboration2023}
 (\bibinfo{year}{EPS-HEP 2023 Conference}),
  \urlprefix\url{https://indico.desy.de/event/34916/contributions/146862/}.

\bibitem[{CMS(ICHEP 2024, Prague)}]{CMS_Collaboration2024}
 (\bibinfo{year}{ICHEP 2024, Prague}),
  \urlprefix\url{https://indico.cern.ch/event/1291157/contributions/5878345/}.

\bibitem[{\citenamefont{Iguro et~al.}(2024)\citenamefont{Iguro, Kitahara, and
  Watanabe}}]{Iguro:2024hyk}
\bibinfo{author}{\bibfnamefont{S.}~\bibnamefont{Iguro}},
  \bibinfo{author}{\bibfnamefont{T.}~\bibnamefont{Kitahara}}, \bibnamefont{and}
  \bibinfo{author}{\bibfnamefont{R.}~\bibnamefont{Watanabe}},
  \bibinfo{journal}{Phys. Rev. D} \textbf{\bibinfo{volume}{110}},
  \bibinfo{pages}{075005} (\bibinfo{year}{2024}), \eprint{2405.06062}.

\bibitem[{\citenamefont{Aaij et~al.}(2020)}]{LHCb:2020cyw}
\bibinfo{author}{\bibfnamefont{R.}~\bibnamefont{Aaij}} \bibnamefont{et~al.}
  (\bibinfo{collaboration}{LHCb}), \bibinfo{journal}{Phys. Rev. D}
  \textbf{\bibinfo{volume}{101}}, \bibinfo{pages}{072004}
  (\bibinfo{year}{2020}), \eprint{2001.03225}.

\bibitem[{\citenamefont{Tanabashi et~al.}(2018)}]{ParticleDataGroup:2018ovx}
\bibinfo{author}{\bibfnamefont{M.}~\bibnamefont{Tanabashi}}
  \bibnamefont{et~al.} (\bibinfo{collaboration}{Particle Data Group}),
  \bibinfo{journal}{Phys. Rev. D} \textbf{\bibinfo{volume}{98}},
  \bibinfo{pages}{030001} (\bibinfo{year}{2018}).

\bibitem[{\citenamefont{Koponen}(2007)}]{Koponen:2007fe}
\bibinfo{author}{\bibfnamefont{J.}~\bibnamefont{Koponen}},
  \bibinfo{journal}{Acta Phys. Polon. B} \textbf{\bibinfo{volume}{38}},
  \bibinfo{pages}{2893} (\bibinfo{year}{2007}), \eprint{hep-lat/0702006}.

\bibitem[{\citenamefont{Li et~al.}(2009)\citenamefont{Li, Lu, and
  Wang}}]{Li:2009wq}
\bibinfo{author}{\bibfnamefont{R.-H.} \bibnamefont{Li}},
  \bibinfo{author}{\bibfnamefont{C.-D.} \bibnamefont{Lu}}, \bibnamefont{and}
  \bibinfo{author}{\bibfnamefont{Y.-M.} \bibnamefont{Wang}},
  \bibinfo{journal}{Phys. Rev. D} \textbf{\bibinfo{volume}{80}},
  \bibinfo{pages}{014005} (\bibinfo{year}{2009}), \eprint{0905.3259}.

\bibitem[{\citenamefont{Bhol}(2014)}]{Bhol_2014}
\bibinfo{author}{\bibfnamefont{A.}~\bibnamefont{Bhol}},
  \bibinfo{journal}{Europhysics Letters} \textbf{\bibinfo{volume}{106}},
  \bibinfo{pages}{31001} (\bibinfo{year}{2014}),
  \urlprefix\url{https://dx.doi.org/10.1209/0295-5075/106/31001}.

\bibitem[{\citenamefont{Bordone et~al.}(2020)\citenamefont{Bordone, Gubernari,
  Huber, Jung, and van Dyk}}]{Bordone:2020gao}
\bibinfo{author}{\bibfnamefont{M.}~\bibnamefont{Bordone}},
  \bibinfo{author}{\bibfnamefont{N.}~\bibnamefont{Gubernari}},
  \bibinfo{author}{\bibfnamefont{T.}~\bibnamefont{Huber}},
  \bibinfo{author}{\bibfnamefont{M.}~\bibnamefont{Jung}}, \bibnamefont{and}
  \bibinfo{author}{\bibfnamefont{D.}~\bibnamefont{van Dyk}},
  \bibinfo{journal}{Eur. Phys. J. C} \textbf{\bibinfo{volume}{80}},
  \bibinfo{pages}{951} (\bibinfo{year}{2020}), \eprint{2007.10338}.

\bibitem[{\citenamefont{Dutta and Rajeev}(2018)}]{Dutta:2018jxz}
\bibinfo{author}{\bibfnamefont{R.}~\bibnamefont{Dutta}} \bibnamefont{and}
  \bibinfo{author}{\bibfnamefont{N.}~\bibnamefont{Rajeev}},
  \bibinfo{journal}{Phys. Rev. D} \textbf{\bibinfo{volume}{97}},
  \bibinfo{pages}{095045} (\bibinfo{year}{2018}), \eprint{1803.03038}.

\bibitem[{\citenamefont{Sahoo and Mohanta}(2019)}]{Sahoo:2019hbu}
\bibinfo{author}{\bibfnamefont{S.}~\bibnamefont{Sahoo}} \bibnamefont{and}
  \bibinfo{author}{\bibfnamefont{R.}~\bibnamefont{Mohanta}}
  (\bibinfo{year}{2019}), \eprint{1910.09269}.

\bibitem[{\citenamefont{Zhang et~al.}(2022)\citenamefont{Zhang, Zhong, Fu,
  Cheng, Zeng, and Wu}}]{Zhang:2022opp}
\bibinfo{author}{\bibfnamefont{Y.}~\bibnamefont{Zhang}},
  \bibinfo{author}{\bibfnamefont{T.}~\bibnamefont{Zhong}},
  \bibinfo{author}{\bibfnamefont{H.-B.} \bibnamefont{Fu}},
  \bibinfo{author}{\bibfnamefont{W.}~\bibnamefont{Cheng}},
  \bibinfo{author}{\bibfnamefont{L.}~\bibnamefont{Zeng}}, \bibnamefont{and}
  \bibinfo{author}{\bibfnamefont{X.-G.} \bibnamefont{Wu}},
  \bibinfo{journal}{Phys. Rev. D} \textbf{\bibinfo{volume}{105}},
  \bibinfo{pages}{096013} (\bibinfo{year}{2022}), \eprint{2202.02730}.

\bibitem[{\citenamefont{Sahoo et~al.}(2021)\citenamefont{Sahoo, Mohanta, and
  Giri}}]{Sahoo:2021wyc}
\bibinfo{author}{\bibfnamefont{S.}~\bibnamefont{Sahoo}},
  \bibinfo{author}{\bibfnamefont{R.}~\bibnamefont{Mohanta}}, \bibnamefont{and}
  \bibinfo{author}{\bibfnamefont{A.~K.} \bibnamefont{Giri}},
  \bibinfo{journal}{Springer Proc. Phys.} \textbf{\bibinfo{volume}{261}},
  \bibinfo{pages}{853} (\bibinfo{year}{2021}).

\bibitem[{\citenamefont{Blossier et~al.}(2022)\citenamefont{Blossier, Cahue,
  Heitger, La~Cesa, Neuendorf, and Zafeiropoulos}}]{Blossier:2021xvl}
\bibinfo{author}{\bibfnamefont{B.}~\bibnamefont{Blossier}},
  \bibinfo{author}{\bibfnamefont{P.-H.} \bibnamefont{Cahue}},
  \bibinfo{author}{\bibfnamefont{J.}~\bibnamefont{Heitger}},
  \bibinfo{author}{\bibfnamefont{S.}~\bibnamefont{La~Cesa}},
  \bibinfo{author}{\bibfnamefont{J.}~\bibnamefont{Neuendorf}},
  \bibnamefont{and}
  \bibinfo{author}{\bibfnamefont{S.}~\bibnamefont{Zafeiropoulos}},
  \bibinfo{journal}{Phys. Rev. D} \textbf{\bibinfo{volume}{105}},
  \bibinfo{pages}{054515} (\bibinfo{year}{2022}), \eprint{2110.10061}.

\bibitem[{\citenamefont{Sahoo and Bhol}(2020)}]{Sahoo:2020wnk}
\bibinfo{author}{\bibfnamefont{S.}~\bibnamefont{Sahoo}} \bibnamefont{and}
  \bibinfo{author}{\bibfnamefont{A.}~\bibnamefont{Bhol}}
  (\bibinfo{year}{2020}), \eprint{2005.12630}.

\bibitem[{\citenamefont{Gubernari et~al.}(2023)\citenamefont{Gubernari,
  Khodjamirian, Mandal, and Mannel}}]{Gubernari:2023rfu}
\bibinfo{author}{\bibfnamefont{N.}~\bibnamefont{Gubernari}},
  \bibinfo{author}{\bibfnamefont{A.}~\bibnamefont{Khodjamirian}},
  \bibinfo{author}{\bibfnamefont{R.}~\bibnamefont{Mandal}}, \bibnamefont{and}
  \bibinfo{author}{\bibfnamefont{T.}~\bibnamefont{Mannel}},
  \bibinfo{journal}{JHEP} \textbf{\bibinfo{volume}{12}}, \bibinfo{pages}{015}
  (\bibinfo{year}{2023}), \eprint{2309.10165}.

\bibitem[{\citenamefont{Rahmani and Ahwazian}(2024)}]{Rahmani:2024pko}
\bibinfo{author}{\bibfnamefont{S.}~\bibnamefont{Rahmani}} \bibnamefont{and}
  \bibinfo{author}{\bibfnamefont{M.}~\bibnamefont{Ahwazian}}
  (\bibinfo{year}{2024}), \eprint{2409.02460}.

\bibitem[{\citenamefont{Sakaki et~al.}(2015)\citenamefont{Sakaki, Tanaka,
  Tayduganov, and Watanabe}}]{Sakaki:2014sea}
\bibinfo{author}{\bibfnamefont{Y.}~\bibnamefont{Sakaki}},
  \bibinfo{author}{\bibfnamefont{M.}~\bibnamefont{Tanaka}},
  \bibinfo{author}{\bibfnamefont{A.}~\bibnamefont{Tayduganov}},
  \bibnamefont{and} \bibinfo{author}{\bibfnamefont{R.}~\bibnamefont{Watanabe}},
  \bibinfo{journal}{Phys. Rev. D} \textbf{\bibinfo{volume}{91}},
  \bibinfo{pages}{114028} (\bibinfo{year}{2015}), \eprint{1412.3761}.

\bibitem[{\citenamefont{Bhattacharya et~al.}(2016)\citenamefont{Bhattacharya,
  Nandi, and Patra}}]{Bhattacharya:2015ida}
\bibinfo{author}{\bibfnamefont{S.}~\bibnamefont{Bhattacharya}},
  \bibinfo{author}{\bibfnamefont{S.}~\bibnamefont{Nandi}}, \bibnamefont{and}
  \bibinfo{author}{\bibfnamefont{S.~K.} \bibnamefont{Patra}},
  \bibinfo{journal}{Phys. Rev. D} \textbf{\bibinfo{volume}{93}},
  \bibinfo{pages}{034011} (\bibinfo{year}{2016}), \eprint{1509.07259}.

\bibitem[{\citenamefont{Celis et~al.}(2017)\citenamefont{Celis, Jung, Li, and
  Pich}}]{Celis:2016azn}
\bibinfo{author}{\bibfnamefont{A.}~\bibnamefont{Celis}},
  \bibinfo{author}{\bibfnamefont{M.}~\bibnamefont{Jung}},
  \bibinfo{author}{\bibfnamefont{X.-Q.} \bibnamefont{Li}}, \bibnamefont{and}
  \bibinfo{author}{\bibfnamefont{A.}~\bibnamefont{Pich}},
  \bibinfo{journal}{Phys. Lett. B} \textbf{\bibinfo{volume}{771}},
  \bibinfo{pages}{168} (\bibinfo{year}{2017}), \eprint{1612.07757}.

\bibitem[{\citenamefont{Tanaka and Watanabe}(2013)}]{Tanaka:2012nw}
\bibinfo{author}{\bibfnamefont{M.}~\bibnamefont{Tanaka}} \bibnamefont{and}
  \bibinfo{author}{\bibfnamefont{R.}~\bibnamefont{Watanabe}},
  \bibinfo{journal}{Phys. Rev. D} \textbf{\bibinfo{volume}{87}},
  \bibinfo{pages}{034028} (\bibinfo{year}{2013}), \eprint{1212.1878}.

\bibitem[{\citenamefont{Duan et~al.}(2025)\citenamefont{Duan, Iguro, Li,
  Watanabe, and Yang}}]{Duan:2024ayo}
\bibinfo{author}{\bibfnamefont{W.-F.} \bibnamefont{Duan}},
  \bibinfo{author}{\bibfnamefont{S.}~\bibnamefont{Iguro}},
  \bibinfo{author}{\bibfnamefont{X.-Q.} \bibnamefont{Li}},
  \bibinfo{author}{\bibfnamefont{R.}~\bibnamefont{Watanabe}}, \bibnamefont{and}
  \bibinfo{author}{\bibfnamefont{Y.-D.} \bibnamefont{Yang}},
  \bibinfo{journal}{JHEP} \textbf{\bibinfo{volume}{07}}, \bibinfo{pages}{166}
  (\bibinfo{year}{2025}), \eprint{2410.21384}.

\bibitem[{\citenamefont{Faroughy et~al.}(2020)\citenamefont{Faroughy, Isidori,
  Wilsch, and Yamamoto}}]{Faroughy:2020ina}
\bibinfo{author}{\bibfnamefont{D.~A.} \bibnamefont{Faroughy}},
  \bibinfo{author}{\bibfnamefont{G.}~\bibnamefont{Isidori}},
  \bibinfo{author}{\bibfnamefont{F.}~\bibnamefont{Wilsch}}, \bibnamefont{and}
  \bibinfo{author}{\bibfnamefont{K.}~\bibnamefont{Yamamoto}},
  \bibinfo{journal}{JHEP} \textbf{\bibinfo{volume}{08}}, \bibinfo{pages}{166}
  (\bibinfo{year}{2020}), \eprint{2005.05366}.

\bibitem[{\citenamefont{Greljo et~al.}(2022)\citenamefont{Greljo, Palavri{\'c},
  and Thomsen}}]{Greljo:2022cah}
\bibinfo{author}{\bibfnamefont{A.}~\bibnamefont{Greljo}},
  \bibinfo{author}{\bibfnamefont{A.}~\bibnamefont{Palavri{\'c}}},
  \bibnamefont{and} \bibinfo{author}{\bibfnamefont{A.~E.}
  \bibnamefont{Thomsen}}, \bibinfo{journal}{JHEP}
  \textbf{\bibinfo{volume}{10}}, \bibinfo{pages}{010} (\bibinfo{year}{2022}),
  \eprint{2203.09561}.

\bibitem[{\citenamefont{Iguro and Watanabe}(2020)}]{Iguro:2020cpg}
\bibinfo{author}{\bibfnamefont{S.}~\bibnamefont{Iguro}} \bibnamefont{and}
  \bibinfo{author}{\bibfnamefont{R.}~\bibnamefont{Watanabe}},
  \bibinfo{journal}{JHEP} \textbf{\bibinfo{volume}{08}}, \bibinfo{pages}{006}
  (\bibinfo{year}{2020}), \eprint{2004.10208}.

\bibitem[{\citenamefont{Abdesselam et~al.}(2019)}]{Belle:2019ewo}
\bibinfo{author}{\bibfnamefont{A.}~\bibnamefont{Abdesselam}}
  \bibnamefont{et~al.} (\bibinfo{collaboration}{Belle}), in
  \emph{\bibinfo{booktitle}{{10th International Workshop on the CKM Unitarity
  Triangle}}} (\bibinfo{year}{2019}), \eprint{1903.03102}.

\bibitem[{\citenamefont{Chen}(2023)}]{Chen:2868260}
\bibinfo{author}{\bibfnamefont{C.}~\bibnamefont{Chen}} (\bibinfo{year}{2023}),
  \urlprefix\url{https://cds.cern.ch/record/2868260}.

\bibitem[{\citenamefont{Aaij et~al.}(2024)}]{LHCb:2023ssl}
\bibinfo{author}{\bibfnamefont{R.}~\bibnamefont{Aaij}} \bibnamefont{et~al.}
  (\bibinfo{collaboration}{LHCb}), \bibinfo{journal}{Phys. Rev. D}
  \textbf{\bibinfo{volume}{110}}, \bibinfo{pages}{092007}
  (\bibinfo{year}{2024}), \eprint{2311.05224}.

\bibitem[{\citenamefont{Alonso et~al.}(2017)\citenamefont{Alonso, Grinstein,
  and Martin~Camalich}}]{Alonso:2016oyd}
\bibinfo{author}{\bibfnamefont{R.}~\bibnamefont{Alonso}},
  \bibinfo{author}{\bibfnamefont{B.}~\bibnamefont{Grinstein}},
  \bibnamefont{and}
  \bibinfo{author}{\bibfnamefont{J.}~\bibnamefont{Martin~Camalich}},
  \bibinfo{journal}{Phys. Rev. Lett.} \textbf{\bibinfo{volume}{118}},
  \bibinfo{pages}{081802} (\bibinfo{year}{2017}), \eprint{1611.06676}.

\bibitem[{\citenamefont{Li et~al.}(2016)\citenamefont{Li, Yang, and
  Zhang}}]{Li:2016vvp}
\bibinfo{author}{\bibfnamefont{X.-Q.} \bibnamefont{Li}},
  \bibinfo{author}{\bibfnamefont{Y.-D.} \bibnamefont{Yang}}, \bibnamefont{and}
  \bibinfo{author}{\bibfnamefont{X.}~\bibnamefont{Zhang}},
  \bibinfo{journal}{JHEP} \textbf{\bibinfo{volume}{08}}, \bibinfo{pages}{054}
  (\bibinfo{year}{2016}), \eprint{1605.09308}.

\bibitem[{\citenamefont{Zuo et~al.}(2024)\citenamefont{Zuo, Fedele, Helsens,
  Hill, Iguro, and Klute}}]{Zuo:2023dzn}
\bibinfo{author}{\bibfnamefont{X.}~\bibnamefont{Zuo}},
  \bibinfo{author}{\bibfnamefont{M.}~\bibnamefont{Fedele}},
  \bibinfo{author}{\bibfnamefont{C.}~\bibnamefont{Helsens}},
  \bibinfo{author}{\bibfnamefont{D.}~\bibnamefont{Hill}},
  \bibinfo{author}{\bibfnamefont{S.}~\bibnamefont{Iguro}}, \bibnamefont{and}
  \bibinfo{author}{\bibfnamefont{M.}~\bibnamefont{Klute}},
  \bibinfo{journal}{Eur. Phys. J. C} \textbf{\bibinfo{volume}{84}},
  \bibinfo{pages}{87} (\bibinfo{year}{2024}), \eprint{2305.02998}.

\bibitem[{\citenamefont{Iguro and Omura}(2023)}]{Iguro:2023prq}
\bibinfo{author}{\bibfnamefont{S.}~\bibnamefont{Iguro}} \bibnamefont{and}
  \bibinfo{author}{\bibfnamefont{Y.}~\bibnamefont{Omura}},
  \bibinfo{journal}{JHEP} \textbf{\bibinfo{volume}{11}}, \bibinfo{pages}{084}
  (\bibinfo{year}{2023}), \eprint{2306.00052}.

\bibitem[{\citenamefont{Altmannshofer et~al.}(2019)}]{Belle-II:2018jsg}
\bibinfo{author}{\bibfnamefont{W.}~\bibnamefont{Altmannshofer}}
  \bibnamefont{et~al.} (\bibinfo{collaboration}{Belle-II}),
  \bibinfo{journal}{PTEP} \textbf{\bibinfo{volume}{2019}},
  \bibinfo{pages}{123C01} (\bibinfo{year}{2019}), \bibinfo{note}{[Erratum: PTEP
  2020, 029201 (2020)]}, \eprint{1808.10567}.

\bibitem[{\citenamefont{Tanaka and Watanabe}(2017)}]{Tanaka:2016ijq}
\bibinfo{author}{\bibfnamefont{M.}~\bibnamefont{Tanaka}} \bibnamefont{and}
  \bibinfo{author}{\bibfnamefont{R.}~\bibnamefont{Watanabe}},
  \bibinfo{journal}{PTEP} \textbf{\bibinfo{volume}{2017}},
  \bibinfo{pages}{013B05} (\bibinfo{year}{2017}), \eprint{1608.05207}.

\bibitem[{\citenamefont{Navas et~al.}(2024)}]{ParticleDataGroup:2024cfk}
\bibinfo{author}{\bibfnamefont{S.}~\bibnamefont{Navas}} \bibnamefont{et~al.}
  (\bibinfo{collaboration}{Particle Data Group}), \bibinfo{journal}{Phys. Rev.
  D} \textbf{\bibinfo{volume}{110}}, \bibinfo{pages}{030001}
  (\bibinfo{year}{2024}).

\bibitem[{\citenamefont{Hamer et~al.}(2016)}]{Belle:2015qal}
\bibinfo{author}{\bibfnamefont{P.}~\bibnamefont{Hamer}} \bibnamefont{et~al.}
  (\bibinfo{collaboration}{Belle}), \bibinfo{journal}{Phys. Rev. D}
  \textbf{\bibinfo{volume}{93}}, \bibinfo{pages}{032007}
  (\bibinfo{year}{2016}), \eprint{1509.06521}.

\bibitem[{\citenamefont{Du et~al.}(2016)\citenamefont{Du, El-Khadra, Gottlieb,
  Kronfeld, Laiho, Lunghi, Van~de Water, and Zhou}}]{Du:2015tda}
\bibinfo{author}{\bibfnamefont{D.}~\bibnamefont{Du}},
  \bibinfo{author}{\bibfnamefont{A.~X.} \bibnamefont{El-Khadra}},
  \bibinfo{author}{\bibfnamefont{S.}~\bibnamefont{Gottlieb}},
  \bibinfo{author}{\bibfnamefont{A.~S.} \bibnamefont{Kronfeld}},
  \bibinfo{author}{\bibfnamefont{J.}~\bibnamefont{Laiho}},
  \bibinfo{author}{\bibfnamefont{E.}~\bibnamefont{Lunghi}},
  \bibinfo{author}{\bibfnamefont{R.~S.} \bibnamefont{Van~de Water}},
  \bibnamefont{and} \bibinfo{author}{\bibfnamefont{R.}~\bibnamefont{Zhou}},
  \bibinfo{journal}{Phys. Rev. D} \textbf{\bibinfo{volume}{93}},
  \bibinfo{pages}{034005} (\bibinfo{year}{2016}), \eprint{1510.02349}.

\bibitem[{\citenamefont{Sakaki et~al.}(2013)\citenamefont{Sakaki, Tanaka,
  Tayduganov, and Watanabe}}]{Sakaki:2013bfa}
\bibinfo{author}{\bibfnamefont{Y.}~\bibnamefont{Sakaki}},
  \bibinfo{author}{\bibfnamefont{M.}~\bibnamefont{Tanaka}},
  \bibinfo{author}{\bibfnamefont{A.}~\bibnamefont{Tayduganov}},
  \bibnamefont{and} \bibinfo{author}{\bibfnamefont{R.}~\bibnamefont{Watanabe}},
  \bibinfo{journal}{Phys. Rev. D} \textbf{\bibinfo{volume}{88}},
  \bibinfo{pages}{094012} (\bibinfo{year}{2013}), \eprint{1309.0301}.

\bibitem[{\citenamefont{McLean et~al.}(2020)\citenamefont{McLean, Davies,
  Koponen, and Lytle}}]{McLean:2019qcx}
\bibinfo{author}{\bibfnamefont{E.}~\bibnamefont{McLean}},
  \bibinfo{author}{\bibfnamefont{C.~T.~H.} \bibnamefont{Davies}},
  \bibinfo{author}{\bibfnamefont{J.}~\bibnamefont{Koponen}}, \bibnamefont{and}
  \bibinfo{author}{\bibfnamefont{A.~T.} \bibnamefont{Lytle}},
  \bibinfo{journal}{Phys. Rev. D} \textbf{\bibinfo{volume}{101}},
  \bibinfo{pages}{074513} (\bibinfo{year}{2020}), \eprint{1906.00701}.

\bibitem[{\citenamefont{Harrison and Davies}(2022)}]{Harrison:2021tol}
\bibinfo{author}{\bibfnamefont{J.}~\bibnamefont{Harrison}} \bibnamefont{and}
  \bibinfo{author}{\bibfnamefont{C.~T.~H.} \bibnamefont{Davies}}
  (\bibinfo{collaboration}{HPQCD}), \bibinfo{journal}{Phys. Rev. D}
  \textbf{\bibinfo{volume}{105}}, \bibinfo{pages}{094506}
  (\bibinfo{year}{2022}), \eprint{2105.11433}.

\bibitem[{\citenamefont{Bourrely et~al.}(2009)\citenamefont{Bourrely, Caprini,
  and Lellouch}}]{Bourrely:2008za}
\bibinfo{author}{\bibfnamefont{C.}~\bibnamefont{Bourrely}},
  \bibinfo{author}{\bibfnamefont{I.}~\bibnamefont{Caprini}}, \bibnamefont{and}
  \bibinfo{author}{\bibfnamefont{L.}~\bibnamefont{Lellouch}},
  \bibinfo{journal}{Phys. Rev. D} \textbf{\bibinfo{volume}{79}},
  \bibinfo{pages}{013008} (\bibinfo{year}{2009}), \bibinfo{note}{[Erratum:
  Phys.Rev.D 82, 099902 (2010)]}, \eprint{0807.2722}.

\bibitem[{\citenamefont{McLean et~al.}(2019)\citenamefont{McLean, Davies,
  Lytle, and Koponen}}]{McLean:2019sds}
\bibinfo{author}{\bibfnamefont{E.}~\bibnamefont{McLean}},
  \bibinfo{author}{\bibfnamefont{C.~T.~H.} \bibnamefont{Davies}},
  \bibinfo{author}{\bibfnamefont{A.~T.} \bibnamefont{Lytle}}, \bibnamefont{and}
  \bibinfo{author}{\bibfnamefont{J.}~\bibnamefont{Koponen}},
  \bibinfo{journal}{Phys. Rev. D} \textbf{\bibinfo{volume}{99}},
  \bibinfo{pages}{114512} (\bibinfo{year}{2019}), \eprint{1904.02046}.

\bibitem[{\citenamefont{Hill}(2006)}]{Hill:2006ub}
\bibinfo{author}{\bibfnamefont{R.~J.} \bibnamefont{Hill}},
  \bibinfo{journal}{eConf} \textbf{\bibinfo{volume}{C060409}},
  \bibinfo{pages}{027} (\bibinfo{year}{2006}), \eprint{hep-ph/0606023}.

\end{thebibliography}

\end{document}